\documentclass[twocolumn,showpacs,amsmath,amssymb,prd]{revtex4}

\usepackage{graphics}% Include figure files
\usepackage{dcolumn}% Align table columns on decimal point
\usepackage{bm}% bold math
\usepackage{epsfig}

\begin{document}

\title{Open charm tomography of cold nuclear matter}

\author{I.~Vitev}
\email{ivitev@lanl.gov}

\author{T.~Goldman}
\email{tgoldman@lanl.gov}

\author{M.~B.~Johnson}
\email{mbjohnson@lanl.gov}

\affiliation{ Los Alamos National Laboratory, 
Theoretical Division and Physics Division, Los Alamos, NM 87545, USA  }

\author{J.~W.~Qiu}
\email{jwq@iastate.edu}

\affiliation{ Department of Physics and Astronomy, 
Iowa State University, Ames, IA 50011, USA }

\begin{abstract}
We study the relative contribution of partonic sub-processes 
to $D$ meson production and $D$ meson-triggered inclusive
di-hadrons to lowest order in perturbative QCD. 
While gluon fusion dominates the creation of large angle 
$D\bar{D}$ pairs, charm on light parton scattering determines 
the yield of single inclusive $D$ mesons. The distinctly 
different non-perturbative fragmentation of $c$ quarks into 
$D$ mesons versus the fragmentation of quarks and gluons into 
light hadrons results in a strong transverse momentum 
dependence of anticharm content of the away-side 
charm-triggered jet.In p+A reactions, we calculate and resum 
the coherent nuclear-enhanced power corrections from the 
final-state partonic scattering in the medium.
We find that single and double inclusive open 
charm production can be suppressed as much as the 
yield of neutral pions from  dynamical high-twist 
shadowing. Effects of energy loss in p+A collisions 
are also investigated phenomenologically and may lead to
significantly weaker transverse momentum dependence of 
the nuclear attenuation.
\end{abstract}
                                               

\maketitle

%%%%%%%%%%%%%%%%%%%%%%%%%%%%%%%%%%%%%%%%%%%%%%%%%%%%%%%%%%%%%%%%%%%

\section{Introduction}

A useful probe of the dense nuclear matter created in collisions 
of heavy nuclei at the relativistic heavy ion collider (RHIC)
is one that is sensitive to dynamical scales and can be both 
cleanly measured experimentally and reliably calculated theoretically.
Because of color confinement, only hard probes, i.e. those with 
large momentum transfers, can be reliably calculated in 
the perturbation theory of Quantum Chromodynamics 
(QCD)~\cite{Collins:gx,Owens:1986mp}. Experimental 
measurements of inclusive particle 
suppression~\cite{Shimomura:2005en,Dunlop:2005xe} are now able to 
test jet quenching theory out to transverse momenta 
of $p_T \sim  20$~GeV~\cite{Vitev:2006uc}.  
On the other hand, typical dynamical scales of the nuclear matter
produced in relativistic heavy ion collisions are on the order of 
hundreds of MeV, which is both much smaller than the scale of a 
hard probe and  non-perturbative. 
Therefore, an ideal probe should be not only ``hard'' but also 
sensitive to this ``soft'' physics.  Open charm production has 
a potential to satisfy these criteria because of the two 
distinctive scales of the open charm meson:
the charm quark mass, $m_c \sim 1.5$~GeV, a relatively hard scale, 
and the binding energy, $\sim M_D-m_c\sim$ hundreds MeV,  
which may be relevant to the fragmentation and dissociation of 
$D$ mesons.

Little quantitative understanding of  medium effects 
exists for heavy-flavor production, due largely to the fact 
that the relevant data is quite sparse at 
present~\cite{Adler:2005xv,Adams:2004fc} and inferred from 
the modified cross sections for non-photonic electrons.  
The experimental situation is rapidly changing, however.
In addition to introducing new mass scales, the advantages 
of using heavy quarks as probes also arise from the fact that 
once formed, heavy mesons live much 
longer ($\sim 10^{-11}$ sec) than the duration of the 
QGP ($\sim 10^{-23}$ sec), travel macroscopic distances 
away from the creation point and can be measured directly. 
Such precision measurements are not only facilitated by the 
boost at forward  rapidities at RHIC and the LHC but also 
impacted by the non-negligible cold nuclear matter 
effects in this kinematic 
region~\cite{Vitev:2003xu,Kharzeev:2003sk,Qiu:2004da,Hwa:2004in}.

With this in mind, we extend the study of 
Refs.~\cite{Qiu:2004da,Qiu:2003vd,Qiu:2004qk} to hadronic collisions   
with heavy-quark final states and consider single- and double-inclusive 
open  charm production. 
Because of the quantitative importance of multiple scattering 
effects for inferring properties of the nuclear matter, the models 
of in-medium interactions should be subjected to experimental 
verification for heavy-flavor production. One may begin to 
gain confidence by applying the theoretical description 
of such effects to heavy ion collisions in simpler situations, 
such as proton-nucleus (p+A) scattering.  
In the this case,  plasma properties are not an issue 
and the main medium effect is, therefore, due to the multiple 
interactions in cold nuclei.
The medium effects that we plan to investigate are also
present in nucleus-nucleus (A+A) reactions during the interaction
time $\tau_{int.} = 2R_A/\gamma \ll \tau_{eq.}  \ll \tau_{QGP}$
and cannot be neglected. 
Here $R_A$ is the nuclear radius, $\gamma$ is the Lorentz gamma
factor of the nucleus, $\tau_{eq.}$ is the equilibration time
and  $\tau_{QGP}$ is the lifetime of the plasma.

Multiple scattering affects the nuclear cross sections in 
various ways depending on how parton propagation is modified 
by the medium as the partons enter into, and emerge from 
the hard scattering. At one extreme, when the scattering from 
the medium is largely  incoherent, the parton modification is
dominated by transverse momentum 
broadening~\cite{Vitev:2003xu}. 
At the other extreme, when the longitudinal momentum transfer 
is small compared to the length of the path of the parton 
as it propagates through the nucleus, the scattering becomes 
coherent which can lead to 
attenuation, or shadowing. The coherent limit is 
described differently in different 
approaches~\cite{Brodsky:2002ue,Frankfurt:2003zd,Kopeliovich:2001hf}. 
In our work its  effects are calculated in terms of 
nuclear-enhanced power 
corrections~\cite{Qiu:2004da,Qiu:2003vd,Qiu:2004qk}. 
We have shown that these could account for a large portion 
or all of nuclear shadowing observed at small Bjorken $x_B$ 
in lepton-nucleus deep inelastic scattering (DIS).
They also contribute to the sizable suppression of 
forward rapidity single- and double-inclusive hadron 
production at RHIC~\cite{Adler:2004eh,Adams:2006uz}. 
As the heavy quark introduces a new mass scale, the 
dependence of the nuclear-size-enhanced power 
corrections on this scale needs to be elucidated.

Inelastic interactions of partons propagating in nuclear matter 
lead to energy loss in both the incoherent~\cite{Gunion:1981qs} 
and the coherent~\cite{Gyulassy:2000er,Gyulassy:2000fs} regimes. 
While  jet quenching effects are best established in A+A 
reactions~\cite{Shimomura:2005en,Dunlop:2005xe,Vitev:2006uc}, 
it was recently shown that energy loss can lead to hadron 
suppression in d+Au collisions at forward rapidities at 
RHIC~\cite{Kopeliovich:2005ym}.
In a preliminary study~\cite{Vitev:2005cg} we 
approximated the effects of such energy loss by a rapidity
shift. We will demonstrate that partonic implementation of energy 
loss, suggested in~\cite{Vitev:2003xu}, gives qualitatively similar 
results but
also allows extraction of the fraction of the energy lost by an 
average parton in cold nuclear matter. Finally, we point out that 
the multi-parton dynamics discussed in this work will become 
even more important in p+A, and consequently A+A, collisions 
at the LHC.

Our paper is organized as follows:
In Section II we outline the theoretical framework for the calculation
of observables in hadronic collisions in which elastic, inelastic and 
coherent multiple scattering effects in  hot and cold nuclear 
matter can be naturally incorporated. In Section III we calculate to 
lowest order (LO) the contribution of the various
partonic sub-processes to single inclusive $D$ meson production and 
$D$ meson-triggered inclusive di-hadron yields in ``elementary'' 
nucleon-nucleon interactions. The light hadron/charm meson fraction 
in the away-side charm-triggered jet is shown to be sensitive to the 
transverse momentum of the associated particle. 
The effect of nuclear-enhanced power corrections and their physical 
interpretation is discussed in Section IV. The suppression of both $D$ 
meson production cross sections and of $D$ meson-triggered correlations 
at forward rapidity at RHIC are also calculated. We discuss in Section V 
the effects of energy loss in cold nuclear matter and demonstrate that
the resulting nuclear suppression has a different transverse
momentum behavior when compared to dynamical high-twist shadowing. 
Our conclusions are given in Section VI. In Appendix A 
we explicitly calculate the LO parton scattering cross sections 
that contribute to single and double inclusive charm production.
Appendix B contains the details of the factorization of the quark and
gluon coherent scattering from the non-perturbative high-twist matrix 
elements. We also demonstrate how to reduce these high-twist 
matrix elements to the scale of power 
corrections per individual nucleon enhanced by the 
nuclear size.

\section{Theoretical Framework}

\begin{figure}[!tb]
\begin{center}
\epsfig{file = 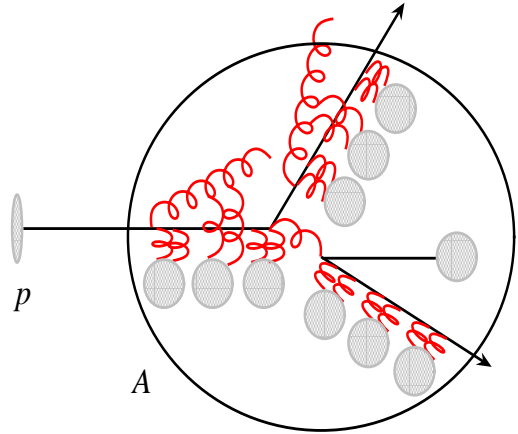,width=3.in,height=2.6in,angle=0}
\vspace*{.1in}
\caption{ Schematic representation of the initial- and 
final-state elastic, inelastic and coherent multiple parton 
scattering that give the nuclear $A^{n/3}$-enhanced corrections
similar to Eq.~(\ref{series}) for proton-nucleus collisions 
in the nuclear rest frame. }
\label{fig-schem}
\end{center} 
\end{figure}

The interpretation of the  
plethora of results from past, present and future 
heavy ion experiments necessitates 
a reliable theoretical framework for the calculation of a
variety of moderate and large transverse momentum observables,
$p_T^2 \gg \Lambda_{\rm QCD}^2$, that can be systematically  
extended to incorporate corrections arising from the many body 
nuclear dynamics. The perturbative QCD factorization 
approach~\cite{Collins:gx}                   
provides such a tool
and expresses the observable hadronic cross sections as  
\begin{eqnarray}
&& \sigma_{\rm hadronic}  =   \sigma_{\rm partonic} 
\left( x_i, z_j;  \mu_r; \mu_{f\, i}, \mu_{d\, j} \right) \;  \nonumber \\
&& \otimes  \; \left\{ \prod_i  \phi_{i/h_i}(x_i, \mu_{f\, i})  \right\} 
 \; \otimes  \;  \left\{ \prod_j D_{h_j/j}(z_j, \mu_{d\, j}) \right\} , \qquad
\label{factorization}
\end{eqnarray}
where  $\otimes$ denotes the standard convolution over 
the internal kinematic variables in the reaction. 
In Eq.~(\ref{factorization}) $\mu_{r},\;\mu_{f\, i}$ and   
$\mu_{d\, j}$ are the renormalization, factorization and 
fragmentation scales, respectively.  $\phi_{i/h_i}(x_i, \mu_{f\, i})$ 
is the distribution function (PDF) of parton ``$i$'' in the hadron  $h_i$ 
and  $D_{h_j/j}(z_j, \mu_{d\, j})$ is the fragmentation or decay 
function (FF) of parton ``$j$'' into hadron $h_j$. Factorization 
not only separates the short- and long-distance QCD dynamics but 
implies universality of the PDFs and FFs and infrared safety of the 
hard scattering partonic cross sections.

\begin{table*}
\begin{tabular}{ll}
\toprule \\
{\bf Type of  scattering}  \ \ \ \  &
{\bf Transverse momentum dependence of the nuclear effect} \\ [3ex]
\toprule \\
{\it Elastic (incoherent)} $\qquad$   & 
Inclusive 1-particle production:  Cronin effect,  enhancement at 
moderate $p_T$, disappears \\ 
&at high $p_T$  \\ 
& Inclusive 2-particle production:
di-jet acoplanarity,  broadening of away-side 
correlations   \\ [3ex] %
{\it Inelastic (final-state)} & Inclusive 1-particle production: 
suppression at all $p_T$, 
weak $p_T$ dependence, amplified 
 near  \\  & the   kinematic bounds \\
 &  Inclusive 2-particle production: suppression of high $p_T$
correlations, enhancement of low \\   %
&   $p_T$ correlations    \\ [3ex]
{\it Inelastic (initial-state)} & Inclusive 1-particle production: 
suppression at all $p_T$, 
weak $p_T$ dependence, amplified 
 near  \\  & the   kinematic bounds \\
 &  Inclusive 2-particle production:  same as inclusive 1 particle
production  \\[3ex]  
{\it Coherent ($t-$channel)}  & Inclusive 1-particle production: 
suppression at low $p_T$, disappears at high $p_T$, pronounced  \\ 
 \hspace*{1.3cm}  {\it ($u-$channel)}    &  $p_T$  dependence \\
& Inclusive 2-particle production: same as inclusive 1 particle
production \\[3ex]
{\it Coherent ($s-$channel)}  & Inclusive 1-particle production: 
enhancement at low $p_T$, 
disappears at high $p_T$, pronounced  \\ 
   &  $p_T$  dependence \\
& Inclusive 2-particle production: same as inclusive 1 particle
production  \\ [3ex]
\botrule
\label{class-1}
\end{tabular}
\caption{Effect of elastic, inelastic and coherent multiple 
scattering on  the transverse momentum dependence of single and 
double inclusive hadron production in the perturbative regime.} 
\end{table*}

In the nuclear environment, corrections in the basic 
perturbative formulas arise from 
the possible nuclear dependence of PDFs and FFs 
as well as  multiple soft parton scattering.
These corrections may be enhanced by the nuclear size and
the density of the quark-gluon plasma (QGP) in the deconfined 
phase. In DIS, cold nuclear matter effects can be systematically 
organized in the framework of the factorization approach 
as follows:                     
\begin{eqnarray}
&& \sigma_{\rm hadron} =
 \nonumber  \\[1ex] 
&& \sigma_0^{(2)} 
\left( 1 + C^{(2)}_1 \alpha_s + C^{(2)}_2  \alpha_s^2 
+   C^{(2)}_3  \alpha_s^3 + \cdots  \right) T^{(2)}
 \nonumber  \\
&& + \frac{ \sigma_0^{(4)} }{Q^2}  
\left( 1 + C^{(4)}_1 \alpha_s + C^{(4)}_2  \alpha_s^2 
+   C^{(4)}_3  \alpha_s^3 + \cdots  \right) T^{(4)} 
  \nonumber  \\
&& + \frac{ \sigma_0^{(6)} }{Q^4} 
 \left( 1 + C^{(6)}_1 \alpha_s + C^{(6)}_2  \alpha_s^2 
+   C^{(6)}_3  \alpha_s^3 + \cdots  \right)  T^{(6)}  
  \nonumber  \\[1ex] 
&& +  \cdots \;\; \quad
\label{series}
\end{eqnarray}     
In Eq.~(\ref{series}) $T^{(i)}$ are the twist ``i'' 
non-perturbative matrix elements. 
$Q^2$ is a typical virtuality, for example
in hadron collisions  we find $Q^2 = - \hat{t},\;  
- \hat{u}$ or $- \hat{s}$.
Standard higher order calculations at leading twist, such as NLO, 
NNLO  correspond to the determination of the 
coefficient functions $C^{(2)}_i$ in  Eq.~(\ref{series}). 
Soft in-medium interactions are controlled by new scales, such 
as the transverse momentum transfer squared per mean free path 
$\mu^2/\lambda$~\cite{Vitev:2006uc,Vitev:2003xu}
and the scale of higher twist per nucleon 
$\xi^2$~\cite{Qiu:2003vd,Qiu:2004qk}. 
While we can write a formal expansion in $\alpha_s$ for 
these quantities, it is usually subject to theoretical 
uncertainties and slow convergence of the 
perturbation series. For the purpose of this paper we consider the
new scales to be properties of the medium and keep track only of 
the nuclear size $A^{1/3}$, or $L$,  enhancement.  
In DIS, high-twist shadowing corresponds to
the identification and resummation of the 
nuclear-size-enhanced higher-twist terms, $\sigma_0^{(i)} T^{(i)}$ in 
Eq.~(\ref{series}). We emphasize that in p+A collisions, 
illustrated in Fig.~\ref{fig-schem}, neither initial- nor 
final-state interactions can be neglected since the coupling strength 
to the medium is qualitatively the same.

From a phenomenological point of view, it is  useful to classify 
the multiple parton interactions based on their experimentally
observable effect on the hadronic spectra and the correlated yield 
of leading di-hadrons from away-side jets, or di-hadron correlations. 
Table~\ref{class-1} provides guidance, consistent with our current 
knowledge and results from this work, to the correspondence between
the nuclear modification to the cross sections measured in A+B 
collisions and the binary scaled p+p result,   
\begin{equation}
 R^{(n)}_{AB}  =  \frac{d\sigma^{h_1 \cdots h_n}_{AB} / 
dy_1 \cdots dy_n d p_{T_1} \cdots  d p_{T_n}} 
{\langle N^{\rm coll}_{AB} \rangle\, d\sigma^{h_1 \cdots h_n}_{NN} / 
dy_1 \cdots dy_n d p_{T_1} \cdots  d p_{T_n}} \; .
\label{multi}
\end{equation}
In Eq.~(\ref{multi})  $\langle N^{\rm coll}_{AB} \rangle$ can be
evaluated from an optical Glauber model.

\section{$D$ meson production and correlations in 
lowest order PQCD}

Nuclear-enhanced power corrections to the 
hadronic cross sections 
are sensitive to the partonic structure of the 
underlying hard scattering event. We first evaluate the 
differential cross sections for open charm production and  
open-charm-triggered di-hadron correlations in the Born 
approximation. Calculations to match the fixed order (FO)
perturbative results to resummed  logarithms of the 
type $ \alpha^n_s (a_k \ln {p_T}/{m_c} )^k $ up to 
next-to-leading log (NLL),  $n=2$,  exist  that treat 
flavor as  ``heavy''~\cite{Cacciari:2005rk}, based
on the assumption $\phi_{c/N}(x,\mu_f)\equiv 0$.
We include the charm contribution
from the nucleon wave-function explicitly since this approach
leads to a faster convergence of the perturbation 
series~\cite{Olness:1997yc}. 
Our goal is to investigate the fractional contribution 
of partonic sub-processes to the differential open charm 
cross section with power corrections in mind. 
We naturally limit our discussion to transverse
momenta larger than the heavy quark mass 
$p_{T_1}, p_{T_2} \geq m_c$.

\subsection{Partonic sub-processes for inclusive $D$ meson 
production and $D$ meson-triggered correlations}

\begin{figure}[t!]
\begin{center}
\epsfig{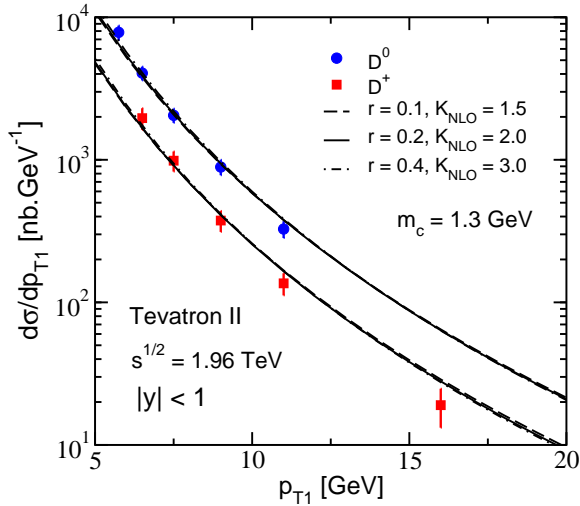}
\caption{ Cross section for $D^0$ and $D^+$
charm meson production to LO in perturbative QCD at
$\sqrt{s}=1.96$~TeV. Three approximately equivalent pairs  
of a phenomenological  $K$-factor and the $D$ meson 
fragmentation parameter $r$  
are shown. }  
\label{fig-TeV}
\end{center} 
\end{figure}

The kinematics that underlies hadronic collisions is 
introduced by  the following example of LO  jet production 
in the collinear factorization approach:
\begin{eqnarray}
\frac{ d\sigma^{j_1(c) j_2(d)}_{NN} }{ dy_c  dy_d d^2p_{T_c}  d^2p_{T_d} } 
&=& 
K_{NLO} \sum_{ab}  \frac{\delta (\Delta \varphi - \pi) 
\delta (p_{T_c}- p_{T_d} ) }{ p_{T_d} } \nonumber \\
&& \hspace*{-1.5cm} 
\times \frac{\phi_{a/N}({x}_a) \phi_{b/N}({x}_b) }{{x}_a{x}_b}  \,
\frac{\alpha_s^2}{{S}^2 }  |\overline {M}_{ab\rightarrow cd}|^2   \;.
\label{basic}
\end{eqnarray}
In  Eq.~(\ref{basic})  $S = (p_A + p_B)^2$  is the 
squared center of mass energy of the hadronic collision,  
 $x_a = p_a^+/{p_A}^+$,   $x_b = p_b^-/{p_B}^-$  are 
the lightcone momentum fractions of the incoming partons and 
$ \phi_{a/N}(x_{a}), \phi_{b/N}(x_{b})$ are the  
distribution function of partons $a,b$ in the nucleon. 
In this work we use the CTEQ6.1 LO  PDFs~\cite{Pumplin:2002vw}. 
In the double collinear approximation the jets are produced 
back-to-back at LO in $\alpha_s$. The running of the strong 
coupling constant  $\alpha_s$ is also taken to lowest order.

The power corrections to single- and double-inclusive hadron production 
are manifest only at small and moderate momentum 
transfers~\cite{Qiu:2003vd,Qiu:2004da},
 where the effect of the heavy quark mass may not be negligible.   
PDFs, based on global QCD analysis, use 
massless evolution kernels  to derive the $Q^2$ dependence of the 
heavy, $c$ and $b$, quark distributions~\cite{Pumplin:2002vw}. 
Systematic corrections of the type $m_q^2 / Q^2$ have so far proven 
difficult to implement. To reconcile the use of parton distributions 
for charm quarks with $m_q = 0$ with a final  state where 
$m_q \neq 0$  we calculate the relevant matrix elements in Appendix A.

With $m_{T_i} = \sqrt{p_{T_i}^2+m_i^2}$, $i =c, d$ we identify the 
lightcone momentum fractions:
\begin{eqnarray}
x_a &=& \frac{1}{\sqrt{S}}( m_{T_c} e^{y_c} + m_{T_d} e^{y_d} ) \;, 
 \\
x_b &=& \frac{1}{\sqrt{S}}( m_{T_c} e^{-y_c} + m_{T_d} e^{-y_d} ) \;.
\label{xis}
\end{eqnarray}
From Eq.~(\ref{basic}), after fragmentation into charm mesons, and  
integration over the away-side jets, we obtain the single 
inclusive  $D^0$, $D^+$ meson production cross section:  
\begin{eqnarray}
\frac{ d\sigma^{D_1}_{NN} }{ dy_1  d^2p_{T_1}  } 
& = & 
K_{NLO} \sum_{abcd} \int_{\cal D}  dy_2 \int_{\cal D}  d z_1 
   \nonumber \\
&& \hspace*{-2cm} \times \frac{1}{z_1^2} D_{D_1/c}(z_1) 
   \frac{\phi_{a/N}({x}_a) \phi_{b/N}({x}_b) }{{x}_a{x}_b} \,
\frac{\alpha_s^2}{{S}^2 }  |\overline {M}_{ab\rightarrow cd}|^2   \;.
\label{single}  
\end{eqnarray}
Here  $${\cal D}=\left\{ (x_1 \in [0,1])  \cap  ( x_2 \in [0,1] ) 
\cap  ( z_1 \in [0,1] )  \right\};$$  
$z_1 = p_{T_1}/p_{T_c}$ and we have chosen the factorization, 
and renormalization scales $\mu_f =\mu_r = 
\sqrt{p_{T_c}^2 + m_c^2} $. 
Only charm quark fragmentation into $D^0, \, D^+$ mesons is considered.

\begin{figure}[t!]
\begin{center}
\epsfig{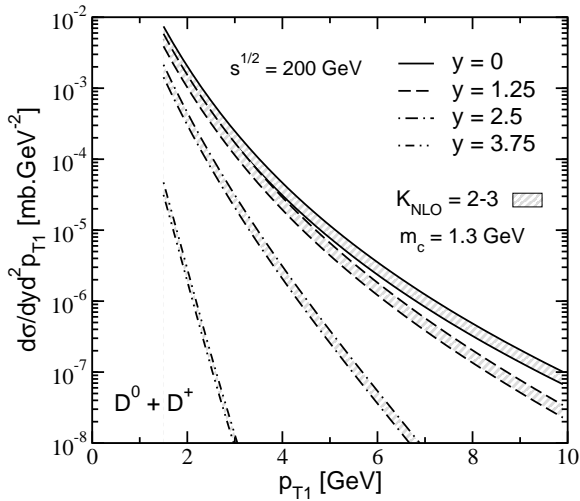}
\caption{ The cross section for $D^0 + D^+$
charm meson production to LO in perturbative QCD at
$\sqrt{s}=200$~GeV in $p+p$ collisions at RHIC with
$K_{NLO} = 2 - 3$ and $r = 0.2$.  The %%
differential cross section is shown in four rapidity bins: 
$y=0, \, 1.25, \,  2.5$ and $3.75$. }  
\label{fig-GeV}
\end{center} 
\end{figure}

In this paper we take  $D_{D/c}(z_1)$  for vector and 
pseudo-scalar mesons from~\cite{Braaten:1994bz}. 
The shape of the heavy quark fragmentation 
function is determined by the value of the non-perturbative 
parameter $r$. Smaller values of $r$ correspond to 
harder fragmentation. The decay of the vector 
states $D^{0*},\; D^{+*}$ into $D^{0},\; D^{+}$ 
with the experimentally measured branching ratios is 
taken into account. 
To study the relation between the phenomenological $K$-factor 
in Eq.~(\ref{single}) and the non-perturbative 
parameter $r$~\cite{Braaten:1994bz} in $D$ meson fragmentation
we compare the calculated $D^0$ and $D^+$ cross sections to the 
Tevatron Run~II $\sqrt{s}= 1.96$~TeV data~\cite{Acosta:2003ax}
in Fig.~\ref{fig-TeV}.  We find that three different combinations  
($K_{NLO}  = 1.5$, $r = 0.1$), ($K_{NLO}  = 2$, $r = 0.2$) and
 ($K_{NLO}  = 3$, $r = 0.4$) yield little difference in 
the  single inclusive charm meson spectrum. We note that 
for the same choice of the parameter $r=0.1$, 
a LO calculation with standard $\phi_{c/N}(x,\mu_f)\neq 0$  gives
open charm cross sections similar to the ones from a NLO 
calculation that treats flavor as ``heavy''.
In this paper we adhere to the more conservative choice 
$r = 0.2 $.  Taking into account an estimated $\sim 50\%$
uncertainty of the $K$-factors at the smaller $\sqrt{s}=200$~GeV, 
we show the  differential $D^0 + D^+$ cross sections in $p+p$ 
in Fig.~\ref{fig-GeV} for several rapidity bins: %
$y = 0, \; 1.25, \; 2.5$ and $3.75$. 
These are the baseline differential distributions 
relative to which the effects of power corrections and 
energy loss can be studied. The uncertainty 
of the choice at RHIC ($K_{NLO}  = 2.5$, $r = 0.2$) cancels in the 
nuclear modification ratios, Eq.~(\ref{multi}).

In hadronic collisions, heavy  quark pair production 
proceeds  via the ``flavor creation'' sub-processes, 
gluon-gluon fusion, $gg \rightarrow c\bar{c}$, and 
light quark-antiquark annihilation, $q\bar{q}\rightarrow c\bar{c}$, 
at lowest order in the strong coupling constant $\alpha_s$.  
Since the gluon density is much larger than the quark and antiquark
distributions at a small parton momentum fraction, $x$, the gluon 
fusion sub-process is believed to dominate
charm production in hadronic collisions at collider energies.
However, for inclusive single charm production, 
a single charm (or anticharm) quark can be created via 
the ``flavor excitation'' sub-processes, $cg\rightarrow cg$ and 
$cq\rightarrow cq$. Although the charm parton distribution is 
small, $\phi_{c/N}(x) \ll \phi_{g/N}(x)$, such processes can 
be amplified by the 
matrix elements $|\overline {M}_{ab\rightarrow cd}|^2$. 
In  Eq.~(\ref{single}), over most of the available
phase space 
\begin{equation}
\left| \frac{\hat{u}}{\hat{t}} \right| \approx 
\left| \frac{\hat{s}}{\hat{t}} \right| = L_1 \gg 1 \;,  
\quad {\rm or} \quad 
\left| \frac{\hat{t}}{\hat{u}} \right| \approx 
\left| \frac{\hat{s}}{\hat{u}} \right| = L_2 \gg 1 \;,  
\label{phasespace} 
\end{equation}
where $L_1, L_2$ are numerically large. 
From the analysis in Appendix A,
we find that the leading power behavior in $L_1$ of 
hard parton scattering for the first case in 
Eq.~(\ref{phasespace}) is 
\begin{eqnarray} 
\label{behavior1}
&&\left\langle | {\cal M}_{gg\rightarrow c\bar{c}} |^{2} \right\rangle 
\sim \frac{1}{6} L_1 \qquad {\rm versus} \\
&&\left\langle | {\cal M}_{cg\rightarrow c{g}} |^{2} \right\rangle 
\sim \frac{8}{9} L_1^2 \;,  \qquad
\left\langle | {\cal M}_{cq\rightarrow c{q}} |^{2} \right\rangle 
\sim 2 L_1^2 \;. \qquad
\label{behavior2}
\end{eqnarray} 
Clearly, there is large, $\propto L_1$, amplification 
of the scattering rate in Eq.~(\ref{behavior2}) relative 
to Eq.~(\ref{behavior1}).

\begin{figure}[t!]
\begin{center}
\hspace*{-.5cm}
\epsfig{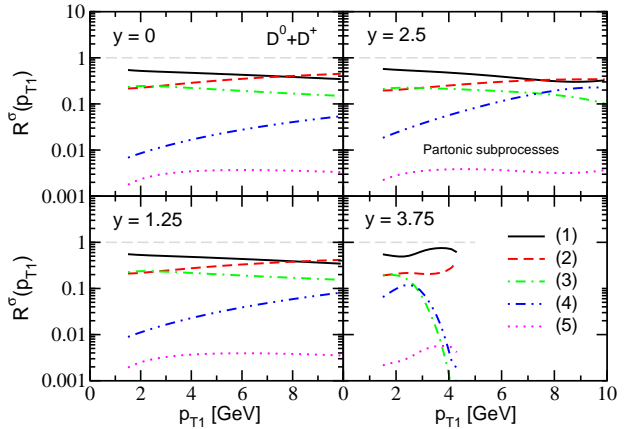}
\caption{ Fractional contribution of partonic sub-processes ($i$) to
moderate and high $p_{T_1}$  $D^0 + D^+$ meson production in 
$\sqrt{s} = 200$~GeV $p+p$  collisions at rapidities 
$y =0,\, 1.25, \, 2.5$ and $3.75$.  We considered (1) $cg \rightarrow cg$
(solid line),
(2) $cq(\bar{q}) \rightarrow cq(\bar{q})$ (dashed line),  
(3) $gg \rightarrow c\bar{c}$ (dot-dashed line), 
(4) $q\bar{q} \rightarrow c\bar{c}$ (double dot-dashed line) and 
(5) $c\bar{c} \rightarrow c\bar{c}$ (dotted line).  Note the dominance %%
(except at the highest values of $p_{T_1}$) of the %%%%%%%% 
$cg \rightarrow cg$ channel.}
\label{fig-SingRat}
\end{center} 
\end{figure}

The fractional weight of the partonic sub-processes to open
charm production can be fully studied through the ratio:
\begin{equation}
R^\sigma(p_{T_1}) = 
\frac{ d\sigma^{D_1}_{ab \rightarrow cd} }
{ dy_1  d^2p_{T_1}  }  \;  \bigg/  
\frac{ d\sigma^{D_1}_{tot} }
{ dy_1  d^2p_{T_1}  } \;  . 
\label{sing-rat}
\end{equation}
Numerical results for the single inclusive charm cross sections, 
Eq.~(\ref{single}), are shown 
in Fig.~\ref{fig-SingRat} at $\sqrt{s}= 200$~GeV. 
We find that the dominant $D$ meson production mechanism is  
the scattering  of the heavy quark from the wavefunction 
of the nucleon (or nucleus) on  gluons, $cg \rightarrow cg$.  
Heavy on light quark scattering, $cq(\bar{q}) \rightarrow 
cq(\bar{q})$, also contributes 
significantly to the cross section.  In these processes the 
charm anti-quarks from the nucleon wavefunction
end up in the fragmentation region near the  beam rapidity.   
While gluon fusion, $gg \rightarrow c\bar{c}$, is  
non-negligible it is also not the main channel for single inclusive
open charm production. We have checked that  $R^\sigma(p_{T_1})$ 
is practically independent of the interplay between $K_{NLO}$
and the hardness $r$ of the $c$ quark fragmentation
into $D$ mesons.

\subsection{Hadron composition of $D$ meson-triggered  
away-side jets}

We now proceed to charm meson-triggered di-hadron 
correlations in away-side  jets, which have not been 
studied previously.  For pairs of hadrons with large
opening angle between them, 
$\Delta \varphi = \varphi_1 - \varphi_2 \sim \pi $, % 
including charm-anticharm and charm-hadron 
correlations,  we find~\cite{Qiu:2004da,Vitev:2005yg}:
\begin{eqnarray}
\!\! \! 
\frac{ d\sigma^{D_1 h_2}_{NN} }{ dy_1  dy_2 d p_{T_1}  d p_{T_2} } 
\! &=&  \! 
K_{NLO} \! \sum_{abcd} \!  
2\pi  \int_{\cal D} 
\frac{d z_1}{z_1}  D_{D_1/c}(z_1)  D_{h_2/d}(z_2)
\nonumber \\
&& \hspace*{-1cm} 
\times  \frac{\phi_{a/N}({x}_a) \phi_{b/N}({x}_b) }{{x}_a{x}_b} \, 
\frac{\alpha_s^2}{{S}^2 }  
|\overline {M}_{ab\rightarrow cd}|^2   \;, 
\label{double} 
\end{eqnarray}
where  
$$\!\!\! {\cal D}=\left\{ ( x_1 \in [0,1] ) \cap ( x_2 \in [0,1] ) 
\cap ( z_1 \in [0,1] \cap ( z_2 \in [0,1] )    \right\}.$$ 
The fragmentation scale for light hadrons in Eq.~(\ref{double}) 
is set to $\mu = p_{T_d}/z_2$  and  $z_2 = z_1 p_{T_2}/ p_{T_1}$.
We use FFs from~\cite{Kniehl:2000fe}
for $\pi^\pm, K^\pm$ and $p(\bar{p})$.  In the calculations that 
follow, we also include charm quark fragmentation into light 
hadrons.  For the charm anti-quark,   
$D_{\bar{D}/{\bar{c}}}(z) = D_{{D}/{{c}}}(z)$.

\begin{figure}[t!]
\begin{center}
\epsfig{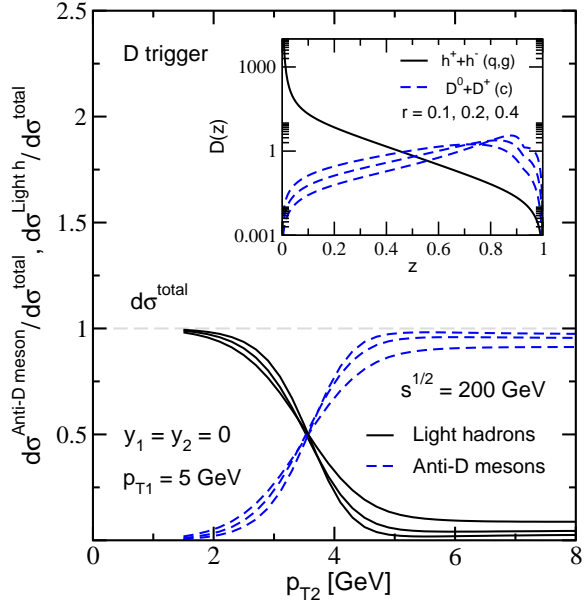}
\vspace*{.1in}
\caption{ Contribution of light hadrons %
and anticharm mesons to the 
{\em fractional production cross section} for 
%%%%%%%%%%%%%%%%
$p_{T_1} = 5$~GeV $D^0+D^+$ meson-triggered away-side
correlations for  $y_1 = y_2 = 0$. The insert
illustrates the difference in the fragmentation of
light quarks and gluons into hadrons versus that for the %%%
$c$ quark into charm mesons. Three different values for the 
non-perturbative fragmentation parameter $r = 0.1,\, 0.2$ and $0.4$
were used.}   
\label{c-trig}
\end{center} 
\end{figure}
The expectation for non-trivial $p_{T_2}$ dependence of  
charm-triggered away-side correlations is based on the 
very different  behavior with respect to  $z = p^{hadron}/p^{parton}$  
of the fragmentation functions of partons into light 
hadrons~\cite{Kniehl:2000fe} compared to those of heavy quarks 
into charm and beauty mesons~\cite{Braaten:1994bz}.  
The insert in Fig.~\ref{c-trig} 
shows that while light hadrons favor %
soft decays of their parent partons, heavy quark fragmentation is 
very hard. For $z \sim 0.6 - 0.9$ there is more than an 
order of magnitude enhancement in the 
decay probabilities $D_{D^0/c}(z)+D_{D^+/c}(z)$ relative to 
$(1/N)\sum_{i=1,N} (D_{\pi^\pm/i} + D_{K^\pm/i} + D_{p(\bar{p})/i})$,
where $i$ runs over the light and charm quarks, antiquarks and 
the gluon.

Triggering on a $D$ meson fixes the momentum of the charm quark 
much more reliably than does triggering on a light 
hadron~\cite{Qiu:2004da,Vitev:2005yg}
in studying away-side correlations. The non-perturbative 
fragmentation therefore controls the  yields of di-hadrons 
versus the associated momentum $p_{T_2}$ and the abundances 
of the different particle species. 
Fig.~\ref{c-trig} shows our prediction for 
the hadronic composition of the ${D}^{0} + D^{+}$-triggered
away-side jet. At transverse momenta significantly smaller 
than the trigger transverse momentum, $p_{T_2} \ll p_{T_1}$, 
the away-side jet is dominated by pions, kaons and protons. 
At transverse momenta  $p_{T_1} \simeq p_{T_2}$, the %
away-side jet is expected to be dominated almost completely 
by $\bar{D}^{0}$ and $D^{-}$ mesons. Since there is little 
sensitivity to the choice of $r$, our prediction is robust 
and can be used as a test of the production mechanism
of heavy quarks.

\begin{figure}[t!]
\begin{center}
\epsfig{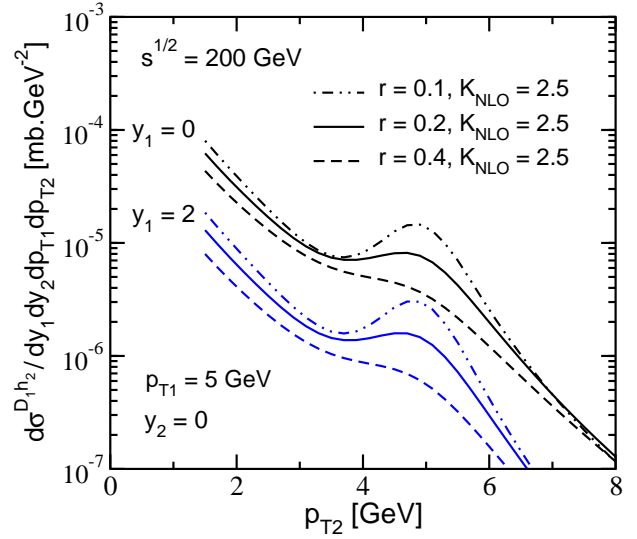}
\vspace*{.1in}
\caption{ The double inclusive hadron production cross section %
at RHIC for a $D^0 + D^+$ trigger at $p_{T_1} = 5$~GeV. 
Three different combinations of $r$ and $K_{NLO}$ are studied.
%%%
Two different rapidity gaps $y_1 - y_2 =0, \, 2$ are shown 
for comparison. The characteristic ``hump'' represents hard 
$\bar{c}$ fragmentation into $\bar{D}$  mesons. } 
\label{c-trig1}
\end{center} 
\end{figure}

Fig.~\ref{c-trig1} shows the behavior of the 
$D$-meson-triggered di-hadron cross section, 
Eq.~({\ref{double}}). The differential spectrum cannot 
be described by a single power law for $p_{T_2} \simeq p_{T_1}$. 
In the case of hard fragmentation, for example $r=0.1$, the 
double inclusive distribution can even be non-monotonic as 
a function of the associated particle momentum  $p_{T_2}$.
The characteristic ``hump'' is associated with the $\bar{c}$ 
fragmentation into $\bar{D}^{0}$ and $D^{-}$. This signature  is  
unique for heavy quark triggers and is seen to  
be independent of rapidity. For pion or unidentified hadron
triggers one expects a smooth power law behavior as a function 
of $p_{T_2}$~\cite{Qiu:2004da,Vitev:2005yg}. We conclude, 
through the comparison of Figs.~\ref{fig-TeV} and~\ref{c-trig1}, 
that charm-triggered correlations provide a much more powerful 
handle on the hardness of $D_{D/c}(z)$ than does 
the single inclusive spectrum.

In summary, even in p+p collisions at RHIC and the LHC,
triggering on charm or beauty mesons can provide 
constraints on the heavy quark production and fragmentation 
mechanisms that are not accessible via the Tevatron single
inclusive data.  Such measurements may already be possible 
at RHIC using electron or muon triggers as proxies for
heavy quarks~\cite{TU,JK}.

\section{Nuclear-enhanced power corrections to 
open charm production}

\begin{figure}[b!]
\includegraphics[width=1.6in]{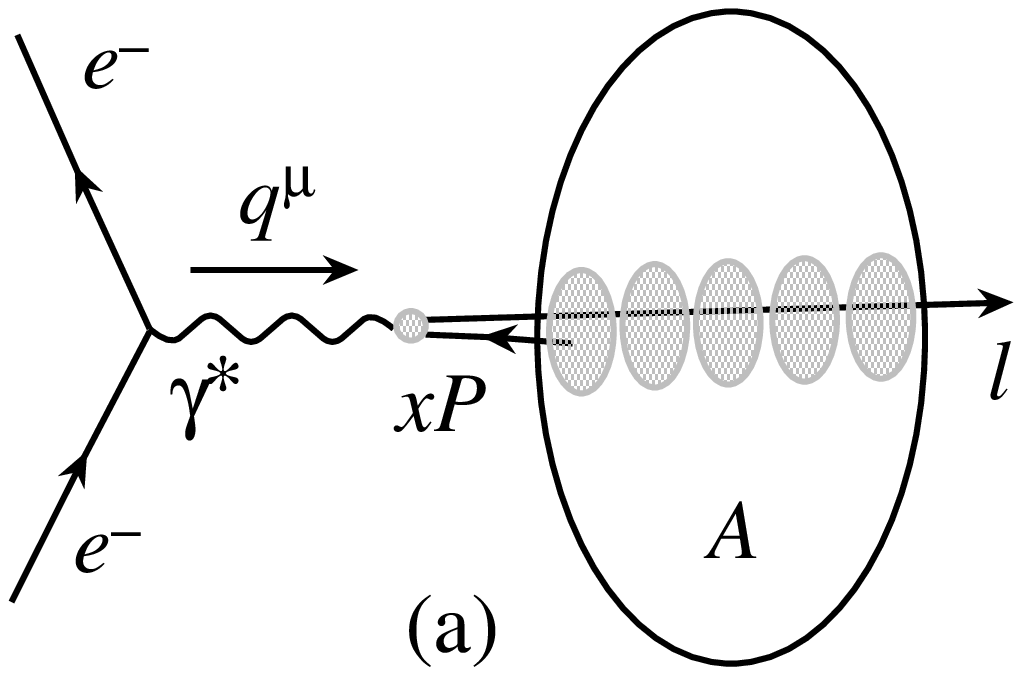}
\includegraphics[width=1.7in]{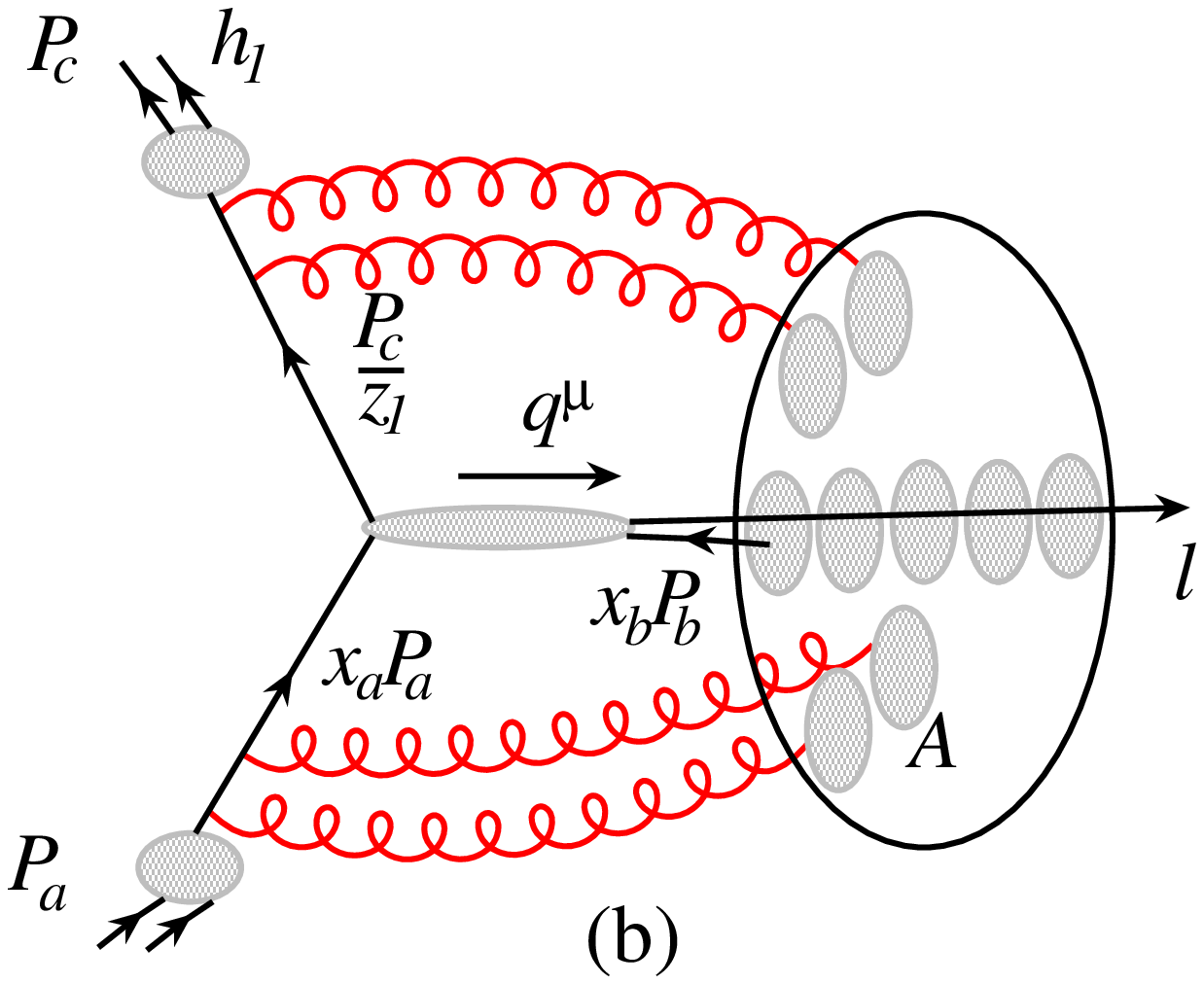}
\caption{ Coherent multiple scattering of the struck parton 
in deeply inelastic scattering~(a) and in proton-nucleus 
collisions~(b). Notice the difference between $p+A$ 
and DIS in terms of the multiple 
initial- and final-state scattering of the incoming and 
outgoing partons (``a'' and ``c''). }
\label{DISandpA}
\end{figure}

In this section we focus on the effects of coherence that 
arise from the multiple soft scattering of the partons that  
participate in the  hard interaction. The processes where 
such effects can be cleanly studied are deeply inelastic scattering
(DIS), illustrated in Fig.~\ref{DISandpA}(a), and Drell-Yan.
If the longitudinal momentum transfer along a fixed direction
becomes small, the probe will interact simultaneously 
with more than one nucleon, due to the uncertainty principle. 
In a frame where the longitudinal 
size of the exchanged virtual meson is given by $l_c = 1/xP$ 
the Lorentz contracted nucleon has longitudinal size 
$2r_0/\gamma = 2r_0/(P/m_N)$.  The critical value of the 
momentum fraction for the onset of  coherence can be easily 
estimated to be  $ x_N = 1 / (2 r_0 m_N) 
\sim 0.1$~\cite{Qiu:2003vd,Qiu:2004qk}. For $x \leq 0.01$, 
the interaction  with all of the nucleons at the same impact 
parameter $b$ is coherent, given
the small transverse size $A_\perp = 1/Q^2$, $Q^2 \geq m_N^2$, 
determined by the virtual probe. 
The multiple final-state interactions with these nucleons are 
mediated via the field of soft gluons and are formally 
suppressed by a power of the squared momentum transfer 
$1/Q^2$ for each additional coherent scattering. 
The resummed contributions form all powers in $1/Q^2$
have the effect of a dynamical generation of a quark or
 gluon {\em mass} as we shall  see below.

In the case of hadronic collisions there is indeed a similarity 
with the DIS dynamics in the final-state rescattering 
of the struck, small $x_b$ parton from the nucleus, as shown 
in  Fig.~\ref{DISandpA}(b). Such nuclear-enhanced power corrections
lead  to suppression of the single and double inclusive       %%
hadron production cross sections  at 
forward rapidity as long as the coherence criterion is 
satisfied~\cite{Qiu:2004da}.

\subsection{Physical interpretation}

The physical interpretation of the resummed nuclear-enhanced 
power corrections is most transparent in the example of 
neutrino-nucleus scattering because of the presence of a 
physical mass scale $M$ in the {\em final} state in some 
of the sub-processes~\cite{Qiu:2004qk}. Even in the absence 
of final-state interactions, the charged current $s$-quark 
to $c$-quark transition, $s\pm W^\pm 
\rightarrow c $, requires 
rescaling of the value of Bjorken $x_B = Q^2/2m_N \nu$  
in the parton distribution functions 
\begin{equation}  
x = x_B \; \rightarrow  \;
x = x_B\left( 1 + \frac{M^2}{Q^2} \right )\;.
\label{heavy}
\end{equation}  
Since $x$ represents the  lightcone momentum fraction carried by 
the struck quark, Eq.~(\ref{heavy}) expresses the conversion
of the larger initial parton energy into the constituent mass of
the heavy quark. In Ref.~\cite{Qiu:2004qk}, using  Dirac 
equations of motion,  we showed that the physical mass and the 
dynamical power corrections commute and therefore the two  effects 
add. The corresponding change in the value of Bjorken $x$ is
\begin{eqnarray}
x = x_B  \rightarrow  \!\!\! && 
x = x_B \left( 1 + \frac{ \xi^2(A^{1/3}-1)}{Q^2}  + 
\frac{M^2}{Q^2} \right) \nonumber \\ 
&& \;\;\; = x_B  \left( 1 +  \frac{ m_{dyn}^2}{Q^2}  + 
\frac{M^2}{Q^2} \right) \nonumber \\ 
&&  \;\;\; = x_B  \left( 1 +  \frac{ M^2}{Q^2} \right)
\left( 1 +   \frac{m_{dyn}^2}{ M^2 +  Q^2 }   \right)\; . \qquad 
\label{heavypq}
\end{eqnarray}
Our resummed higher-twist corrections rescale the 
momentum fraction, $x_B$, in Eq.~(\ref{heavypq}) in a manner
identical to the modification arising from the constituent 
heavy quark mass, Eq.~(\ref{heavy}). We identify 
$m^2_{dyn} = \xi^2 (A^{1/3}-1)$ 
with the dynamical mass~\cite{Qiu:2004da,Qiu:2003vd,Qiu:2004qk} 
generated by propagation of the struck parton  through the
nuclear background chromo-magnetic 
field.  We remark that this effect, which we calculated in QCD, 
is analogous to the generation of dynamical mass for 
electrons propagating in a strong electro-magnetic
field~\cite{Itoh:1998xk}.  For $x_B < 0.1$, due to coherence, 
$m^2_{dyn}$  has to be incorporated 
in the underlying  kinematics of the hard scattering, 
Eq.~(\ref{heavypq}).  For values of  $x_B > 0.1$ rescattering 
of the struck parton still occurs but the vector meson 
exchange is localized to one nucleon. 
The elastic final-state interactions vanish by 
unitarity upon integrating over the full phase space for 
the hadronic fragments, $X$.

\begin{figure}[t!]
\psfig{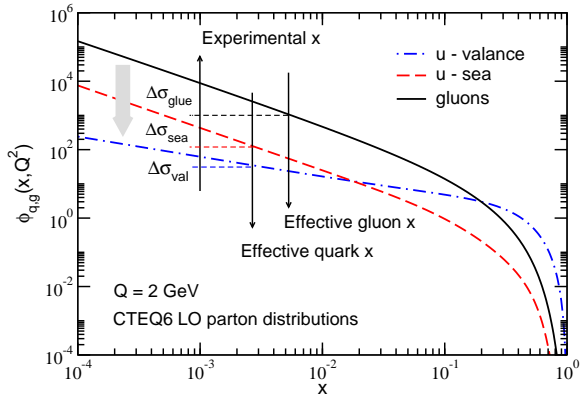}
\caption{ Effects of rescaling of the momentum fraction
$x$ illustrated on the example of the lowest order CTEQ6 parton     %%
distribution functions for valence and sea u-quarks. Gluons  %
rescatter in the final-state to lowest order only in hadronic 
collisions and are included for completeness. The effective 
parton distributions or partonic flux are reduced after the
rescaling of the momentum fraction.
We used  $Q^2 = 4$~GeV$^2$ and the large rescaling in $x$ is for 
purposes of illustration 
only. } 
\label{mechanism}
\end{figure}

The mechanism behind the reduction of the deeply inelastic
inclusive scattering cross section and the differential hadron 
production cross sections is illustrated  schematically 
in Fig.~\ref{mechanism}. 
PDFs are falling functions of the momentum fraction $x$: 
\begin{equation}   
\begin{array}{ll}
\phi_{\rm v}(x)  \propto  x^{-1/2 + \delta_v}   \;,  \qquad  
              &   \qquad {\rm  valence \; quarks}\; ,    \\[2ex] 
\phi_{\rm s}(x) \propto  x^{-1 + \delta_s}   \;,  \qquad  
              &    \qquad {\rm  sea \; quarks} \; ,     \\[2ex] 
\phi_{\rm g}(x) \propto  x^{-1 + \delta_g}   \; ,  \qquad  
              & \qquad {\rm  gluons} \; , 
\end{array}
\label{pdf-s-x}
\end{equation}
where $|\delta_{v,s,g}| \ll 1$.
The evaluation of the parton  distribution function at an 
effectively larger value of $x$  leads to a 
reduction of the cross section,       
$-\Delta \sigma$~\cite{Qiu:2003vd,Qiu:2004qk}. 
From Eq.~(\ref{pdf-s-x}) it follows that the same 
dynamically generated parton mass for quarks apparently 
produces a different reduction in the sea and valence 
contributions  to the DIS cross sections, 
$ - \Delta \sigma_{\rm sea} \simeq   - 2\Delta \sigma_{\rm valence}  $.  
If a gluon  scatters coherently in the final state,     
the generated dynamical mass  is $C_A/C_F = 9/4$ larger than 
the dynamical mass for a quark due to the larger average squared 
color charge. Consequently, the observable  ``shadowing'' of gluons  
is substantially larger than that for sea quarks, despite the 
very similar low-$x$ behavior of the PDFs in Eq.~(\ref{pdf-s-x}),
$ - \Delta \sigma_{\rm gluon} \simeq -  9/4 \Delta \sigma_{\rm sea} $.  
We can, therefore,  derive a hierarchy of  dynamical high-twist
 shadowing, 
\begin{eqnarray}
&& S(x,Q^2) = \left|\; \frac{  - \Delta \sigma (x,Q^2) }
          {  \sigma (x, Q^2) } \; \right| \nonumber \\[1ex]
&& S_v (x,Q^2) < S_s(x,Q^2) < S_g(x,Q^2)  \;,     
\label{hier}
\end{eqnarray}
for cross sections with multiple scattering dominated by
valence quarks, sea quarks or gluons, respectively.

\subsection{Power corrections in p+A reactions with 
            heavy quark final states}

A detailed derivation of nuclear-enhanced power corrections in 
proton(deuteron)-nucleus collisions is given in Appendix B. 
In this subsection we 
summarize  the main results. Unlike structure functions in DIS, 
hadronic collisions involve a more complicated momentum flow 
even in LO tree level diagrams. This requires isolating 
the $x_b$ dependence of the integrand in Eqs.~(\ref{single})  
and (\ref{double}) in the function 
\begin{equation}
F_{ab\rightarrow cd}(x_b) = \frac{\phi_{b/N}(x_b) 
|M_{ab\rightarrow cd}|^2}{x_b}  \;.
\label{xb-dep}
\end{equation}  
Since we use the standard on-shell hard scattering matrix elements 
$|M_{ab\rightarrow cd}|^2$ reproduced in Appendix A we obtain
an upper limit on the dynamical high-twist shadowing. 
The scale of high-twist corrections on an {\em individual} 
nucleon $\xi^2=0.09 - 0.12$~GeV$^2$ is fixed by the world's
data on deep inelastic scattering on a wide range 
of nuclear targets~\cite{Qiu:2003vd}, including the 
shadowing in the structure functions, $F_2$, and the 
enhancement in the longitudinal structure function, 
$F_L$, relative to the result at leading twist. This is the same 
scale that was used to predict the $Q^2$-dependent suppression  
of the small $x$ structure functions in neutrino 
scattering on iron (Fe) targets measured by the NuTeV 
collaboration~\cite{Tzanov:2005kr}. Since $\xi^2$ is 
constrained from final-state quark scattering, we account
for the representation of 
parton $d$ in hadronic reactions by the color factor $C_d$
on a process-by-process basis. The values are
$C_{q(\bar{q})}=1$ and $C_g=C_A/C_F=9/4$ for a quark 
(antiquark) and a gluon, respectively.

The forward rapidity particle production in p(d)+A collisions is 
analogous to DIS, and resumming the coherent scattering
with multiple nucleons is equivalent to a shift of the momentum
fraction of the active parton from the nucleus which leads to a net
suppression of the cross sections. The hard scattering scale 
is given here by $\hat{t}=q^2=(x_ap_A-p_c/z_1)^2$,
and for Eqs.~(\ref{single}) and (\ref{double}) we derive
the following nuclear modification:
\begin{eqnarray}
 F_{ab\rightarrow cd}(x_b) && \Rightarrow  \nonumber \\  
&&  \hspace*{-1cm}
 F_{ab\rightarrow cd}\left(x_b \left[ 1+ C_d
\frac{\xi^2(A^{1/3}-1) }{-\hat{t}+m_d^2} 
\right] \right)  \;.
\label{resumt}
\end{eqnarray} 
We denote by $A$ the atomic mass number of the nucleus
even though it is labeled as hadron ``$B$'' or ``$b$''.
The $\hat{t}$-dependence of this shift in Eq.~(\ref{resumt}) 
indicates  that the attenuation increases in the 
forward rapidity region but disappears at large transverse 
momenta. We note that the  effect of the mass of the 
final-state parton, for example a charm quark, 
in Eq.~(\ref{resumt}) is precisely the same as in the 
DIS case~\cite{Qiu:2004qk}, Eq.~(\ref{heavypq}).

 \begin{figure}[t!]
\begin{center} 
\psfig{file = 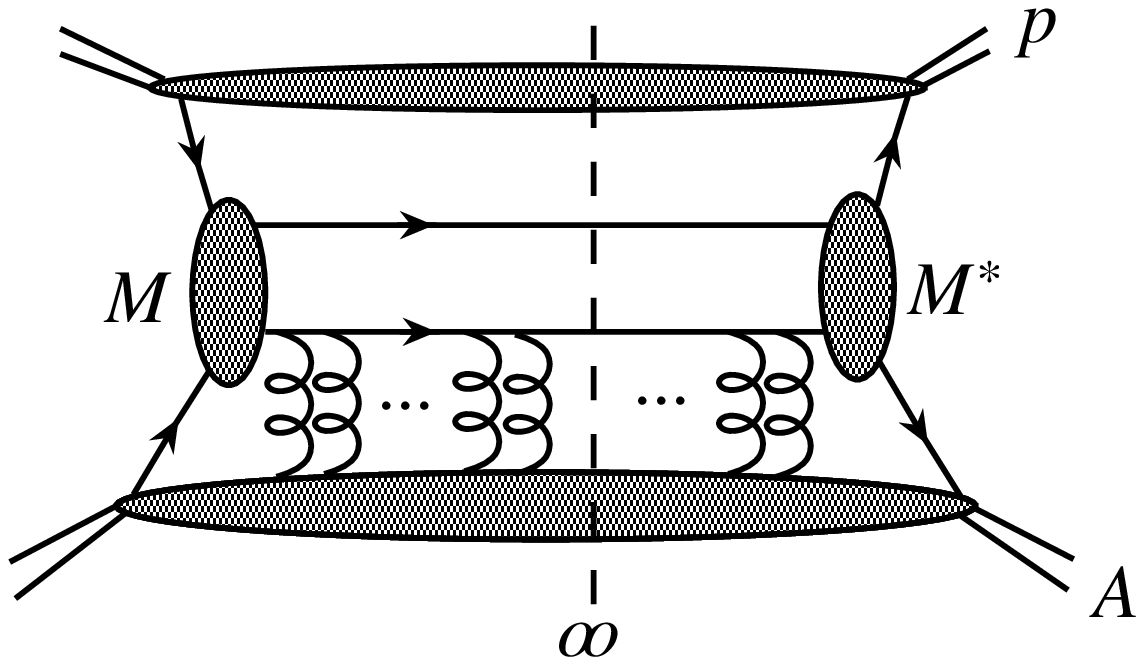,width=1.8in,height=1.2in,angle=0}
\psfig{file = 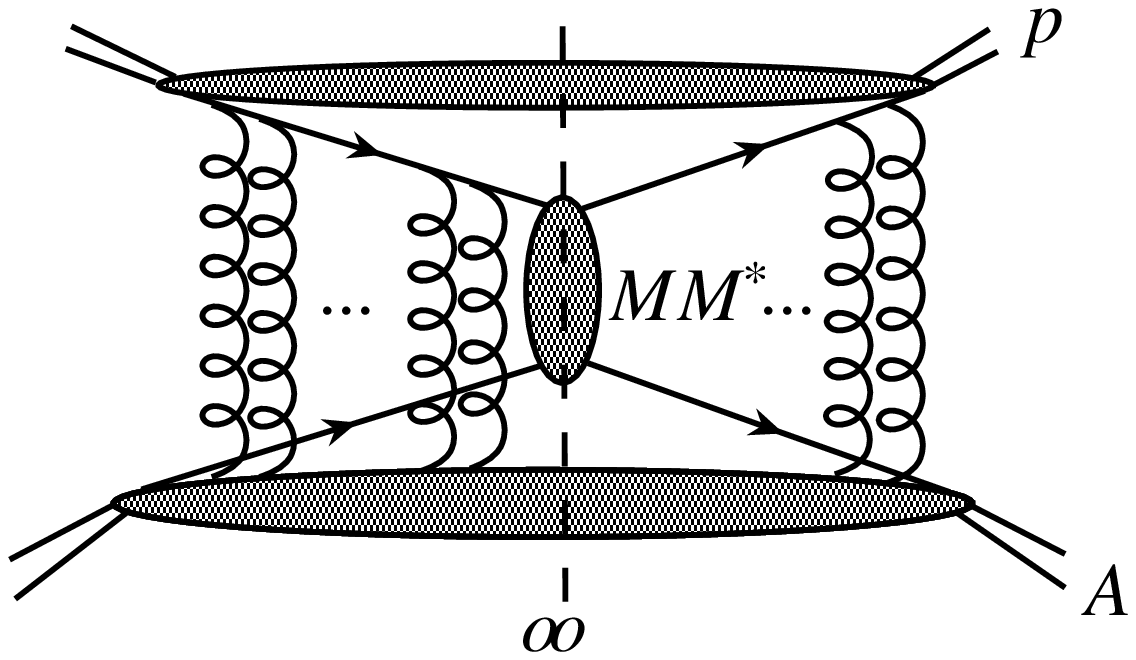,width=1.7in,height=1.2in,angle=0}
\psfig{file = 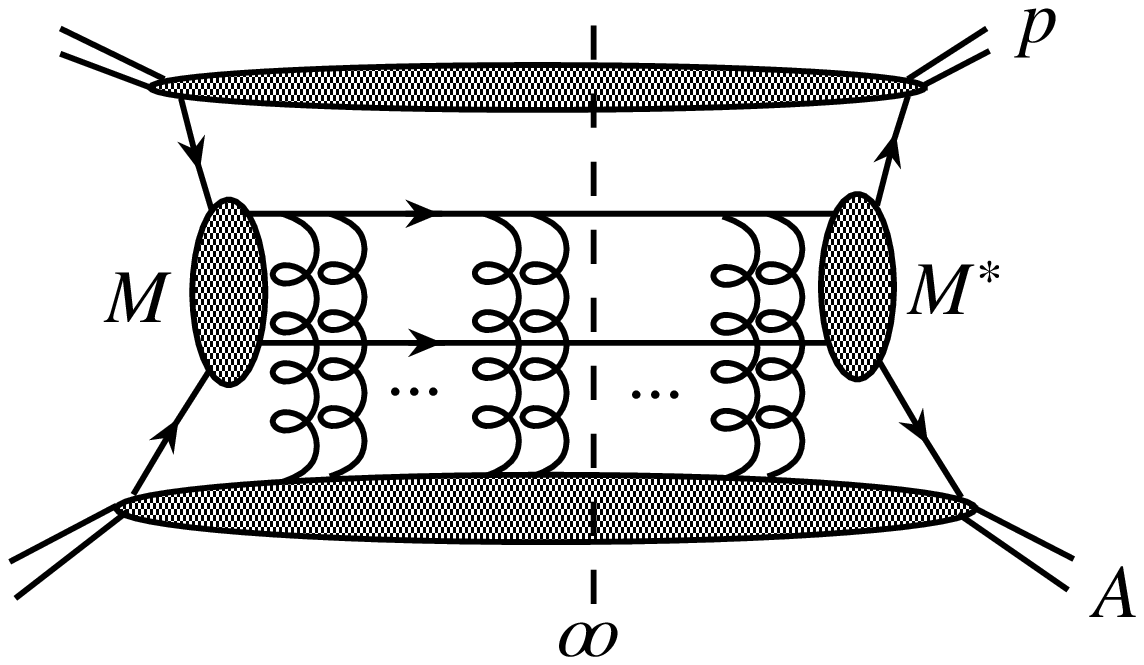,width=1.6in,height=1.05in,angle=0}
\caption{Multiple coherent scattering of the incoming and outgoing 
partons in p+A reactions in the $t$-, $s$- and $u$-channels, 
respectively.} 
\label{allchanels}
\end{center} 
\end{figure}

While we have focused in detail on the  power corrections 
in the $t$-channel, it should be noted that the coherent initial- 
and final-state interactions of the incoming parton from the 
proton, see the bottom diagrams in Fig.~(\ref{allchanels}),  
can also be calculated. The results that we present here are 
analogous to Eq.~(\ref{resumt}), 
\begin{eqnarray}
\label{resums}
 F_{ab\rightarrow cd}(x_b) && \Rightarrow  \nonumber \\  
&&  \hspace*{-1cm} F_{ab\rightarrow cd}\left(x_b \left[ 1+ C_a
\frac{\xi^2(A^{1/3}-1)}{-\hat{s}}  \right]\right)  \;, \\
 F_{ab\rightarrow cd}(x_b) && \Rightarrow  \nonumber \\
&&  \hspace*{-1cm}  F_{ab\rightarrow cd}\left(x_b \left[ 1+ C_c
\frac{\xi^2(A^{1/3}-1) }{-\hat{u}+m_c^2}  \right]\right)  \;. 
\label{resumu}
\end{eqnarray}
In Eqs.~(\ref{resums}) and (\ref{resumu}) $C_a$ and $C_c$ are
the numerical coefficients analogous to $C_d$  corresponding
to partons $a$ and $c$, respectively.  By direct comparison  
of the nuclear-$A^{1/3}$-enhanced power corrections in 
Eqs.~(\ref{resumt}), (\ref{resums}) and (\ref{resumu}) 
it is easy to see that in the forward rapidity region  
$|\hat{t}| \ll |\hat{s}|,  |\hat{u}|$ only DIS-like 
dynamical shadowing corrections are important.

On the other hand, in processes with no strong final-state 
interactions, such as the Drell-Yan, coherent multiple
initial-state scattering can give only {\em  enhancement} 
of the cross section, as shown in Eq.~(\ref{resums}), because 
of the rescaling of $x_b$ to smaller versus larger values. 
Such result is also easily understandable since an incoming 
parton with large dynamically generated mass prior to the 
hard scatter has to carry a smaller fraction of the nucleon 
momentum in order to produce particles in the final state with 
fixed momentum. Thus, initial-state multiple soft 
interactions without energy loss always lead to a Cronin 
effect~\cite{Vitev:2003xu} independently of whether 
they are treated as coherent or not.

The derivations given
here are critical to elucidating the dynamical origin
of nuclear effects in high energy hadronic reactions.  
We have shown through explicit calculation that such
effects are process dependent and may change their 
sign. Therefore, these are {\em not} factorizable as
a part of the PDFs and FFs.

The cross section for single and double inclusive hadron 
production can be calculated from Eqs.~(\ref{single}) and
(\ref{double})  via the substitutions Eqs.~(\ref{resumt}),
(\ref{resums}) and (\ref{resumu}) in minimum bias reactions. 
We recall that the $A^{1/3}$-enhanced scale of higher twist, 
$\xi^2 = 0.12$~GeV$^2$, has the minimum bias geometry included. 
For specific centrality classes we account for the 
varying nuclear thickness along the line of parton 
propagation as follows:
\begin{equation}
A^{1/3} \rightarrow A^{1/3} \frac{T_A(b)}
{T_A(b_{\rm min.bias})} \; ,
\label{central}
\end{equation}
where $T_A(b) = \int_{-\infty}^{\infty}\rho_A(z,b)dz$.

\subsection{Numerical results}

%%%%%%%%%%%%%%%%%%%%%%%%%%%%%%%%%%%%%%%%%%%%%%%%%%%%%%%%%%%%%%%%

\begin{figure}[t!]
\begin{center} 
\hspace*{-0.1in} 
\psfig{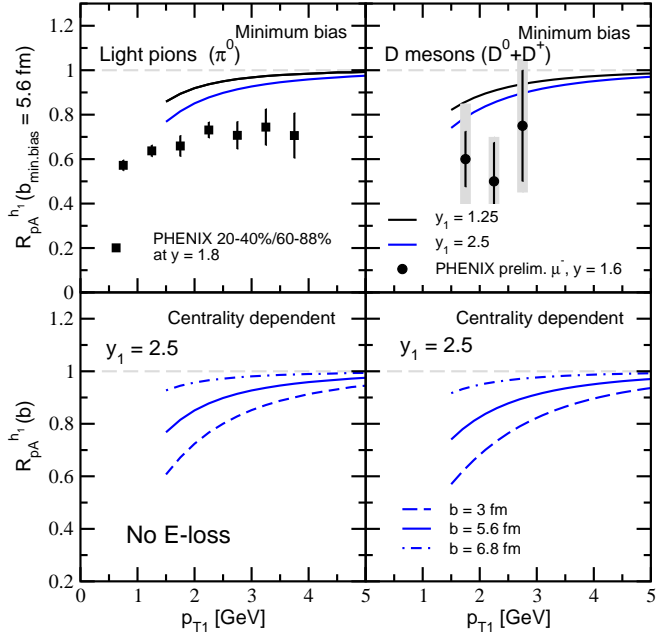}
%%MBJ Start
\caption{Top: suppression of the single inclusive 
hadron production rates in d+Au reactions versus $p_T$ for 
rapidities $y_1 = 1.25$ (smaller attenuation) and $y_1 = 2.5$ 
(larger attenuation).  Bottom: impact parameter 
dependence of the calculated  nuclear modification for central, 
$b=3$~fm, minimum bias, $b_{\rm min.bias}=5.6$~fm and peripheral,
$b=6.8$~fm, collisions. The scale of higher twist 
per nucleon $\xi^2 = 0.12$~GeV$^2$.  
Data is from PHENIX~\cite{Adler:2004eh,Wang:2006de}. 
Theory does not include energy loss. }  
\label{fig-inc-pc}
\end{center} 
\end{figure}

%%%%%%%%%%%%%%%%%%%%%%%%%%%%%%%%%%%%%%%%%%%%%%%%%%%%%%%%%%%%%%%%

To illustrate the similarities and differences between 
massless and massive final-state partons, we carry out a 
comparative study of the effect of power corrections on single
and double inclusive $\pi^0$ and $D$ meson production. 
The left panels of Fig.~\ref{fig-inc-pc} show the 
suppression of the low and moderate $p_{T_1}$ neutral 
pion cross section relative to the binary scaled p+p 
result, Eq.~(\ref{multi}), from high-twist shadowing. 
The nuclear modification factor is shown for two different 
forward rapidities, $y_1 = 1.25, 2.5$, in 
$\sqrt{s}=200$~GeV d+Au collisions at RHIC. At transverse momenta
$p_{T_1} = 1.5$~GeV the suppression can be as large as 25\%
but disappears toward higher $p_{T_1}$ due to the power
law nature of the effect. $R_{dAu}(p_{T_1})$ also shows 
non-zero rapidity dependence with the effect increasing 
at forward rapidities.  For comparison, we show the 
PHENIX measurement
of the nuclear modification of muons coming from the decay 
of charged hadrons in mid-central d+Au reactions at $y_1 = 1.8$ 
is  shown~\cite{Adler:2004eh}. 
Coherent power corrections cannot fully  account for
the nuclear suppression measured by experiment. The discrepancy 
becomes larger at higher transverse momenta. The bottom left panel 
of Fig.~\ref{fig-inc-pc} shows the centrality dependence of
coherent power corrections for light pions. 
From Eq.~(\ref{central}), we see that in central
collisions the larger nuclear thickness generates 
correspondingly larger dynamical parton mass leading to 
larger suppression of the cross sections.

The right panels of Fig.~\ref{fig-inc-pc} show the calculated
suppression of $D^0+D^+$ mesons (and equivalently, 
$\bar{D}^0+D^-$ mesons),  in 
deuteron-gold collisions at RHIC.  It should be noted that
the nuclear modification is very similar to or slightly larger 
than that for $\pi^0$. The reason for this similarity is 
that in both  cases the dominant channel of hadron 
production is via quarks
scattering on gluons, which have the same (large) color singlet 
coupling to the medium, $C_d = 9/4$ in Eq.~(\ref{resumt}). 
In addition, the typical momentum fraction, $z_1$, which enters in 
the determination of the hard scale, $\hat{t} \propto 1 / z_1$, 
is slightly  larger for the $D$ mesons. Preliminary PHENIX data
on the modification of $\mu^-$'s coming  from
the decay of heavy flavor at $y = 1.6$ in cold nuclear matter is 
provided as a reference~\cite{Wang:2006de}. Although 
the error bars are large, $D$ mesons seem to be suppressed  as
much as light hadrons. This attenuation effect is larger 
than predicted by power corrections alone.

\begin{figure}[t!]
\begin{center} 
\psfig{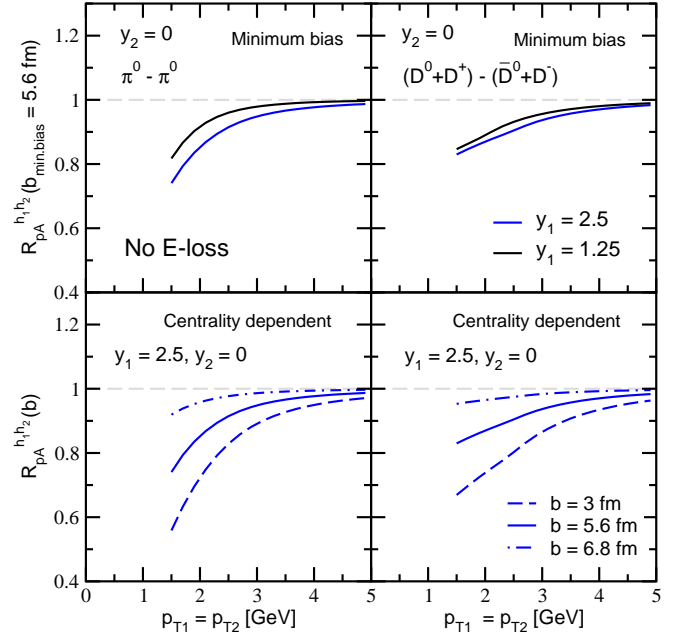}
%%MBJ Start
\caption{Top: suppression of double inclusive 
hadron production rates in d+Au reactions versus $p_{T}$ for 
trigger and associated hadrons of the same transverse 
momentum but different rapidities: $y_1 = 1.25$ (smaller 
attenuation) and $y_1 =  2.5$ (larger attenuation) for 
the trigger and $y_2=0$ for the associated hadron. Bottom: impact 
parameter dependence of the calculated 
nuclear modification for central, $b=3$~fm,
minimum bias, $b_{\rm min.bias}=5.6$~fm and peripheral,
$b=6.8$~fm, collisions. The scale of higher twist per 
nucleon $\xi^2 = 0.12$~GeV$^2$. Theory does not 
include energy loss. }  
\label{fig-corr-pc}
\end{center} 
\end{figure}

Our results for the dynamical shadowing effects in inclusive
two-particle measurements at large angle $\Delta \phi \sim \pi$ 
are given in Fig.~\ref{fig-corr-pc}. We have chosen 
$p_{T_1} = p_{T_2}$ and the same rapidity of the trigger 
hadron, $y_1$, as in Fig.~\ref{fig-inc-pc}. 
The associated hadron was 
chosen at rapidity $y_2 = 0$ to facilitate the study of 
di-jets with a large rapidity gap.  The left panels of 
Fig.~\ref{fig-corr-pc} show that $\pi^0 - \pi^0$ modification
in the nuclear medium follows qualitatively the suppression of
single inclusive pions  as shown in Fig.~\ref{fig-inc-pc}.  
The right panels of
Fig.~\ref{fig-corr-pc} illustrate the effects of  coherent 
power corrections on $(D^0+D^+) - (\bar{D}^0+D^-)$ pair 
production. In this case  we have a heavy $\bar{c}$ quark 
 scattering coherently in the final state. 
At $p_{T_1} = p_{T_2} = 1.5$~GeV $\sim m_c$, the effect of 
the heavy quark mass on the power corrections that we 
derived in Appendix B is noticeable, see Eq.~(\ref{resumt}).
At higher transverse momenta the mass effect is not relevant. 
Naively, we would expect that $\bar{D} D$ pair production 
would  be less suppressed due to the quark versus 
the gluon color singlet coupling to the medium, $C_d = 1$
versus 9/4. However, in 
light hadron pair production the momentum fractions 
$z_1, z_2$ are significantly smaller~\cite{Vitev:2004kd} 
than in single inclusive measurements and $D$ meson 
production. Therefore the relevant hard scale, $\hat{t}$, 
in $\bar{D}D$ pair production is smaller which amplifies 
the effect of the {\em dynamical} heavy quark mass.

%%%%%%%%%%%%%%%%%%%%%%%%%%%%%%%%%%%%%%%%%%%%%%%%%%%%%%%%%%%%%%%%

\subsection{ Discussion }

The nuclear-size-enhanced power corrections that enter this work 
are the result of a tree-level QCD calculation and represent the 
effect of multiple coherent scatterings of the outgoing partons 
with the gluon field of the nucleus. The strength of 
this interaction is specified by 
a universal parameter, $\xi^2$, which is related 
to the limiting soft 
gluon distribution of a nucleon in the nucleus. Once $\xi^2$ is 
specified, the effect  of coherent scattering for all partons, 
heavy and light quarks as well as gluons, is completely determined. 
In particular, when the coherence length, defined as the 
inverse of the longitudinal  momentum transfer, becomes comparable 
to the nuclear size, these  power corrections acting on partons 
in the final state suppress the hadron cross sections.

In Ref. \cite{Kopeliovich:2001hf}, the effects of coherence are 
determined by multiple scattering, and this approach is therefore, 
in principle, quite similar to ours. The following points of 
difference should, however, be noted. Using a Green-function 
description of quark-gluon Fock states in combination with the 
Glauber-Gribov formulation of nuclear shadowing, it is shown there
 how multiple diffractive scattering of gluons suppresses the 
gluon distribution in a nucleus when 
the coherence length in gluon-dominated processes 
begins to exceed the inter-particle spacing, thus providing an 
estimate of gluon shadowing.  The scale of the suppression 
is fixed by, among other things, the
triple-pomeron coupling $G_{3P}=3$~mb/GeV$^2$.  The small triple
 pomeron coupling suggests a rather confined gluon cloud of 
valence quarks, and this weakens gluon shadowing and delays
 its onset.  This gluon shadowing is also  "leading twist", 
i.e., weakly dependent on $Q^2$,  vanishing logarithmically 
as $Q^2 \rightarrow \infty$.

\section{Energy loss in  cold nuclear matter}

In Fig.~\ref{fig-schem} and the discussion in the previous 
section we clarified  that, in the presence of a nucleus, 
final-state rescattering of the struck small $x$ parton with 
its remnants exhausts the similarity 
between hadronic collisions, such as p+A and A+A,  and
DIS~\cite{Qiu:2004da}. 
In $\ell + A$ (DIS), the multiple interactions of the        
incoming  leptons are suppressed by powers of
$\alpha_{em}/\alpha_s$ relative to the struck parton 
scattering.  In contrast, in p+A and A+A the initial- 
and final-state scattering of the incoming  quarks 
and gluons are equally strong. Nuclear modification, 
in particular jet energy loss associated with the 
suppression of particle production, 
cannot be neglected. In a model of Sudakov suppression, 
Ref.~\cite{Kopeliovich:2005ym}, it was shown that the effect 
is amplified at forward Feynman $x$, $x_F \rightarrow 1$.

\subsection{ Evidence for energy loss in cold nuclear matter}

\begin{figure}[!t]
\psfig{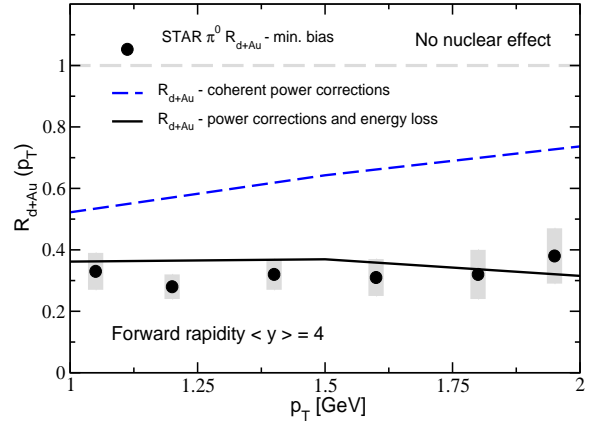}
\caption{ Suppression of $\pi^0$ production at $y=4 $ for 
d+Au collisions at $\sqrt{s_{NN}}=200$~GeV at 
RHIC~\cite{Adams:2006uz}. Calculations of dynamical 
shadowing with and without cold nuclear 
matter energy loss are shown. } 
\label{star}
\end{figure}

To illustrate the importance of energy loss, we study the 
nuclear modification of hadron production over a large range
of center of mass energies and momentum fractions $x_b$. 
In one extreme,  at $\sqrt{s}=200$~GeV d+Au collisions at RHIC,
the STAR collaboration observed a factor of 3 suppression 
of $\pi^0$ production relative to the binary collision scaled
p+p result~\cite{Adams:2006uz}. 
Figure~\ref{star} shows that dynamical high-twist
shadowing, constrained by the DIS data down to values of 
Bjorken $x_B \sim 10^{-4}$, underpredicts the nuclear 
attenuation  by a factor of 2.

Next, we implement the presently known incoherent 
limit of energy loss for on-shell partons first derived 
in~\cite{Gunion:1981qs}. The double differential 
gluon intensity spectrum per scattering reads:   
\begin{equation}
\frac{\omega dN^{(1)}_g}{d \omega d^2{\bf k}}
\propto  \frac{\alpha_s}{\pi^2}
\frac{{\bf q}_1^2} {{\bf k}^2({\bf k} - {\bf q}_1)^2} \; ,
\label{BGrad}
\end{equation}
where ${\bf k}$ is the transverse momentum of the radiative gluon, 
${\bf q}_1$ is the transverse momentum transfer from the medium and
$\omega$ is the gluon energy. The two characteristics  of this 
radiative energy loss that we use are $\Delta E \propto E$ 
and $\Delta E \propto L / \lambda$ in the case of multiple
interactions. Jet quenching before the 
partonic hard scatter implies that in p+A and A+A reactions 
quarks and gluons carry larger fraction $x_a$ of the nucleon
momentum than in p+p collisions~\cite{Vitev:2003xu}:
\begin{eqnarray} 
\label{i-shift}
&& \epsilon = \frac{\Delta E}{E} \propto \frac{L}{\lambda} 
= \kappa A^{1/3}  \;,  \\  
&&  \phi_{q,g/N}(x,\mu_f) \rightarrow 
\phi_{q,g/N}\left( \frac{x}{1-\epsilon}, \mu_f \right) \;.
\label{i-eloss}
\end{eqnarray}
Fitting Eq.~(\ref{i-shift}) to the data, 
we find $\kappa = 0.0175$ 
for minimum bias reactions implying that an average parton 
loses $\sim 10\%$
of its energy in a large nucleus such as Au or Pb.
The variation of  $ \kappa $ with centrality in the Bertsch-Gunion
model is given by the 
nuclear thickness, $T_A(b)$, in a manner similar to the variation 
of the nuclear-enhanced power corrections. Figure.~\ref{star} 
shows that such energy loss implementation gives good 
quantitative agreement with the forward rapidity suppression 
measured by STAR.

\begin{figure}[!t]
\psfig{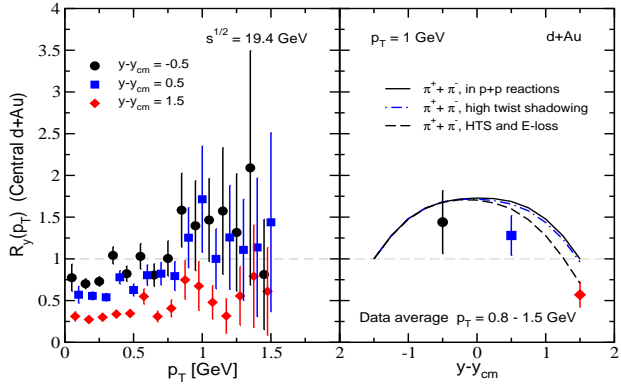}
\caption{ Left panel: nuclear modification at three different
rapidities $y-y_{cm}$ in $\sqrt{s_{NN}}=19.4$~GeV  d+Au 
collisions. Data is from NA35~\cite{Alber:1997sn}.
Right panel: theoretical calculations of charged pion 
production in p+p and d+Au collisions with high-twist shadowing
(HTS) and energy loss compared to the experiment.}
\label{na35}
\end{figure}

At the other extreme, the CERN NA35 fixed target experiment 
with  $y_{cm}=3$   measured hadron production in d+Au reactions at 
$\sqrt{s_{NN}}=19.4$~GeV~\cite{Alber:1997sn}. Although there is a 
factor  $\sim 100$ difference in the values of $x_b$, the STAR 
data were taken   
at much more forward $y$ than the NA35 data,  the same 
rapidity   asymmetry is observed as at RHIC.  
We take $R_y$,  the ratio of hadron spectra 
in different rapidity bins defined as
\begin{equation} 
R_y(p_T) = \frac{d \sigma^{h}(y)} {dy d^2p_T} 
\bigg/    
\frac{d \sigma^{h}(y_{\rm base})} {dy d^2p_T}   \;,
\label{Ry}
\end{equation}  
as a measure of cold nuclear matter effects. 
The left panel of Fig.~\ref{na35} shows this forward 
rapidity suppression of negative hadrons 
relative to the baseline $y-y_{cm} \approx -1.5$ production 
cross section. The right panel of Fig.~\ref{na35} shows 
the deviation of the data in $p_T = 0.8 - 1.5$~GeV interval 
from the  symmetric p+p distribution around $y_{\rm cm}$. 
Dynamical shadowing calculations~\cite{Qiu:2004da} give 
$ < 5\%$ effect at this energy in this rapidity range. 
Leading-twist shadowing parameterizations are also not compatible
with the asymmetry in the data~\cite{Frankfurt:2003zd}. 
Conversely, the implementation of energy loss in cold 
nuclear matter leads to much larger suppression 
at forward rapidity and significant improvement 
in the theoretical description of the data in Fig.~\ref{na35}. 
We conclude that such low energy p+A 
measurements~\cite{Alber:1997sn}  remove the ambiguity 
of what nuclear effects dominate the forward rapidity 
suppression in heavy ion collisions.

\subsection{ Application of energy loss to D meson production
in proton-nucleus collisions }

\begin{figure}[t!]
\begin{center} 
\hspace*{-0.1in} 
\psfig{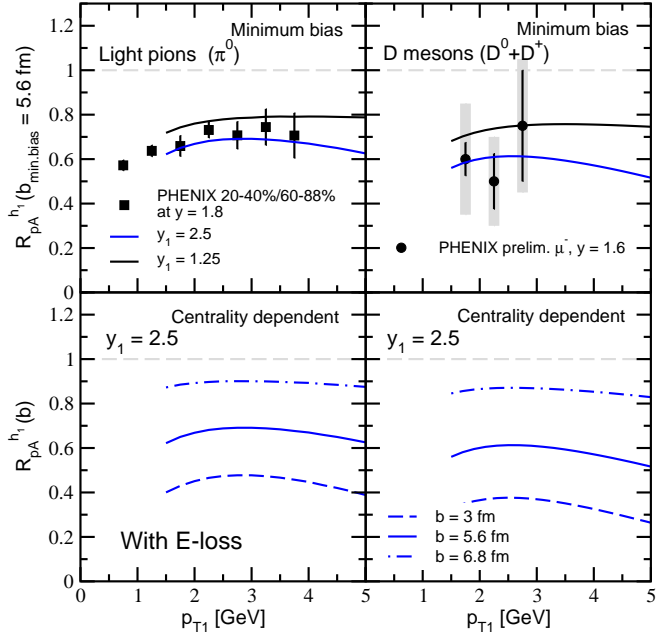}
\caption{ Nuclear modification of single inclusive $\pi^0$ 
(left panels)
and $D^0 + D^+$  mesons (right panels) in d+Au collisions at RHIC. 
The same rapidities and centralities as in  Fig.~\ref{fig-inc-pc} 
are shown. 
Energy loss is taken into account in the theoretical calculations.}
\label{fig-inc-pc-el}
\end{center} 
\end{figure}

\begin{figure}[t!]
\begin{center} 
\psfig{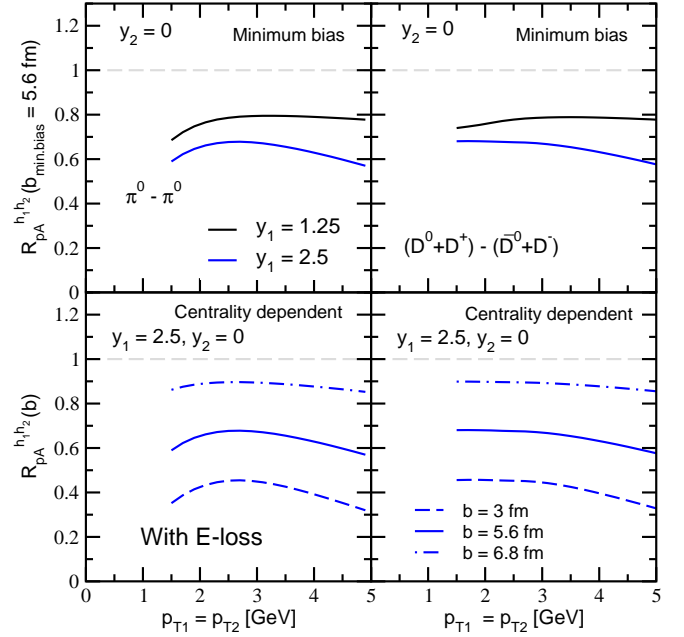}
%%MBJ Start
\caption{  Nuclear modification of $\pi^0 - \pi^0$ di-hadron 
correlations (left panels)
and $(D^0 + D^+) -  (\bar{D}^0 + D^-)$  mesons correlations 
(right panels) in d+Au collisions at RHIC. 
The same rapidities and centralities as in  
Fig.~\ref{fig-corr-pc} are shown. Energy loss is taken into 
account in the theoretical calculations.  }
\label{fig-corr-pc-el}
\end{center} 
\end{figure}

Having investigated the largest dynamic range
of measurements in proton(deuteron)-nucleus reactions 
accessible to perturbative QCD calculations, we return to 
$D$ meson production and correlations at RHIC. We use
the same fractional energy loss $\epsilon = \Delta E / E$ 
as in Figs.~\ref{star} and~\ref{na35}. The forward rapidity 
boost $E = m_T \cosh y_1 \gg m_c$ allows one to neglect the
effect of the charm quark mass on $\Delta E$.  
Single inclusive $\pi^0$ and $D^0+D^+$ suppression at rapidities 
$y_1 = 1.25, \, 2.5$ and minimum bias, central and 
peripheral d+Au collisions at RHIC are shown in 
Fig.~\ref{fig-inc-pc-el}. In this case very good agreement
between the QCD theory incorporating cold nuclear matter 
effects and the PHENIX measurement of muons coming from the
decay of light hadrons~\cite{Adler:2004eh} is achieved.    
We find that the magnitude of the $D$ meson suppression 
is similar to that of pions. It is also comparable with
the first forward rapidity results on heavy quark nuclear 
modification in d+Au reactions at RHIC~\cite{Wang:2006de}.  
In Fig.~\ref{fig-corr-pc-el} we also show the result 
of our calculations for the 
$\pi^0 - \pi^0$ and  $(D^0 + D^+) - (\bar{D}^0 + D^-)$  
mesons pair production at large angles. The same rapidity
gaps, $ \Delta y = y_1 - y_2 $, and centralities as in 
Fig.~\ref{fig-corr-pc} were used. Nuclear modification of
inclusive two particle production  is seen to follow 
closely that of single inclusive hadrons.

There are similarities  between the calculations that
only include resummed nuclear-enhanced power corrections,  
Figs.~\ref{fig-inc-pc} and~\ref{fig-corr-pc}, and 
the calculations that do not neglect energy loss in cold 
nuclear matter, Figs.~\ref{fig-inc-pc-el} and~\ref{fig-corr-pc-el}. 
Both effects are  generated  through multiple 
parton scattering in the medium and lead to suppression of 
the rate of hard scattering. In both cases single inclusive
particle production and large angle di-hadron correlations 
are similarly attenuated. Like all nuclear many body effects 
these increase with the centrality of the collision. 
The difference in the resulting nuclear modification 
is that high-twist shadowing arises from
the coherent final-state scattering of the struck small $x_b$ 
parton of the nucleus and disappears as a function of the
transverse momentum. The energy loss considered here 
arises from the initial-state inelastic scattering of the 
incoming large $x_a$ parton from the proton(deuteron) 
and leads to a suppression which is much more $p_T$ 
independent. By comparing our results for inclusive one-
and inclusive two-particle production we see that
\begin{equation} 
  \frac{  R^{(2)}_{d+A}(p_{T_1} = p_{T_2} ) } 
 { R^{(1)}_{d+A}(p_{T_1}) } 
= \frac{ \frac{d\sigma^{h_1 h_2}_{dAu} 
/ dy_1 dy_2 d p_{T_1} d p_{T_2} } 
{\langle N^{\rm coll}_{dAu} \rangle\, d\sigma^{h_1  h_2}_{NN} / 
dy_1 dy_2 d p_{T_1} d p_{T_2} }  }
{  \frac{d\sigma^{h_1 }_{dAu} / dy_1 d p_{T_1} } 
{\langle N^{\rm coll}_{dAu} \rangle\, d\sigma^{h_1 }_{NN} / 
dy_1 d p_{T_1} }   }
\approx 1 \; . 
\label{IAA}
\end{equation} 
Deviation from unity may arise due to specific choices of 
$p_{T_1}, p_{T_2}$ and improved theoretical calculations of 
the parton color charge, system size and jet energy 
dependence of $\Delta E$ in Eq.~(\ref{i-shift}). Both 
calculations presented in this paper, are consistent 
with the lack of {\em significant} modification of the 
per trigger yield in d+Au reactions relative to 
p+p~\cite{Adler:2006hi}.

%%%%%%%%%%%%%%%%%%%%%%%%%%%%%%%%%%%%%%%%%%%%%%%%%%%%%%%%%%%%%%%%

\section{Conclusions}

In this paper we have systematically studied inclusive $D$ meson 
production and $D$ meson-triggered di-hadron yields in p+p 
and p(d)+A reactions.

At collider energies, we have found that while $D\bar{D}$ 
pair production is
dominated by heavy  quark  flavor creation processes,
such as gluon-gluon fusion $gg\rightarrow c\bar{c}$ and 
light quark-antiquark annihilation $q\bar{q}\rightarrow c\bar{c}$,
single inclusive $D$ (and $\bar{D}$) meson production is 
controlled by radiatively generated charm quarks in  
the nucleon wave function~\cite{Olness:1997yc,Pumplin:2002vw}
scattering on light quarks and gluons $cg\rightarrow cg$, 
$cq(\bar{q})\rightarrow cq(\bar{q})$. Our results for the 
mechanism of $D$ meson production are readily testable at 
RHIC and LHC via charm quark triggered two particle 
measurements~\cite{TU,JK}. It is a robust prediction of 
this approach that at $p_{T_2} < p_{T_1}$ the $D$ meson 
triggered away-side jet is dominated  by  hadrons 
from the fragmentation of light quarks and
gluons. Only at transverse momenta $p_{T_2} \geq p_{T_1}$
is the associated yield  found to be dominated by $\bar{D}$ 
mesons. Such experimental measurements will  be 
able to shed light on the partonic processes responsible 
for $D$ meson production and also constrain the 
hardness of the non-perturbative heavy quark fragmentation 
functions~\cite{Braaten:1994bz}. Our results are also 
suggestive that inclusive heavy meson cross sections 
may be larger than the ones obtained in perturbative
expansions that always treat $c$ and $b$ 
quarks as ``heavy''~\cite{Cacciari:2005rk}.

%%%%%%%%%%%%%%%%%%%%%%%%%%%%%%%%%%%%%%%%%%%%%%%%%%%%%%%%%%%%%%%%

In p(d)+A reactions we calculated and resummed the 
nuclear-enhanced coherent power 
corrections~\cite{Qiu:2003vd,Qiu:2004qk} to single 
inclusive $D$ meson production and $D$ meson-triggered di-hadron 
yields. Improving upon our previous work~\cite{Qiu:2004da}, we 
established here the effect of the heavy quark mass on the power 
corrections and found it to be relatively small for charm quarks. 
We also extended the study of coherent multiple scattering to
all $s$-, $t$- and $u$- channels.  While forward rapidity at 
RHIC is dominated by DIS-like $t$-channel final-state 
interactions other kinematic domains and other particle production 
processes, such as the Drell-Yan, are controlled by different
channels. In this paper, by direct calculation, we demonstrated 
that {\em initial}-state $s$-channel multiple interactions lead 
to an enhancement in the cross sections in clear contrast to 
final-state suppression.  We have thus further corroborated the 
conclusion that enhancement/suppression effects in reactions 
with nuclei are dynamically generated in the 
collision~\cite{Vitev:2006uc,Vitev:2003xu,Qiu:2004da,Gyulassy:2000er,Gyulassy:2000fs}. 
We find that they are not universal (are process dependent) and cannot 
be separated in the parton distribution or fragmentation 
functions.

Numerically, we found that dynamical high-twist shadowing
can suppress single inclusive $D$ meson production by as much 
as it does light hadrons. Similar results were obtained 
for di-hadron
production with light and heavy quark triggers in the 
rapidity range $y_1 = 1 - 4$ studied at RHIC. The nuclear 
modification at forward rapidity and small $p_T \sim 1$~GeV 
can be substantial, $\sim 30 - 50 \%$. At high $p_T$, however,
it disappears quickly $\propto 1/p_T^2$ and cannot fully 
explain the nuclear suppression measured in d+Au reactions by 
PHENIX~\cite{Adler:2004eh,Wang:2006de} and STAR~\cite{Adams:2006uz}. 
Our calculations are fully constrained by existing DIS 
data~\cite{Qiu:2003vd} and provide an upper
limit on nuclear modification from dynamical high-twist 
shadowing. Deviations between the calculations and the 
existing data are indicative of inelastic scattering in
nuclei~\cite{Kopeliovich:2005ym}.

To determine the additional 
dynamical effects that lead to forward-backward hadron
production asymmetry in proton-nucleus collisions  
in a model-independent way,
we investigated low energy d+Au collisions~\cite{Alber:1997sn} 
where coherent scattering does not play a role. 
Suppression of forward rapidity hadrons, similar to that 
observed at RHIC, is found and is consistent
with jet energy loss in cold nuclear matter. We have shown that,
based on initial-state partonic implementation of energy 
loss~\cite{Vitev:2003xu} similar to the established
incorporation of final-state $\Delta E$~\cite{Vitev:2006uc}, 
the forward rapidity  suppression can be 
described over the widest dynamic range in p(d)+A
collisions accessible to perturbative 
QCD analysis. Specifically, we implemented radiative 
energy loss of the type derived by Bertsch and 
Gunion~\cite{Gunion:1981qs} and found 
that if the average parton loses $\sim 10\%$ of its energy 
in minimum bias reactions with nuclei such as Au and Pb, 
this can explain the suppression of forward hadron 
production from SPS to RHIC. Inclusion of cold nuclear matter 
energy loss together with coherent power corrections  
results in suppression of both single inclusive hadron 
production and hadron triggered 
inclusive two particle yields (including $D$ mesons and 
$D\bar{D}$ pair production)  which are fully consistent 
with the forward rapidity hadron attenuation  measurements 
at RHIC~\cite{Adler:2004eh,Wang:2006de,Adams:2006uz}. 

The implication of such energy loss for extracting the 
scale of momentum transfer in cold nuclear matter merits 
further investigation. Of particular interest are 
analytic solutions for {\em initial}-state gluon 
bremsstrahlung, including coherence  effects to all 
orders in the correlation between multiple scattering 
centers. In the future,  These can be studied 
numerically and implemented in the pQCD formalism with a degree of 
sophistication that will match the well-developed 
final-state energy  loss phenomenology~\cite{Vitev:2006uc}.

In summary, we have presented the first perturbative QCD calculation
on heavy meson triggered di-hadron yields and showed how it 
can provide information on the $D$ meson production 
mechanism and the non-perturbative fragmentation of heavy quarks.
We identified the nuclear effects that dominate the cold 
nuclear matter attenuation of $D$ meson and light hadron 
production at forward rapidities, which has drawn considerable
interest in its own right. This work also provides
the baseline for precision QGP tomography using heavy quarks as 
probes of the plasma away from midrapidity where new experimental 
capabilities at RHIC and the LHC are expected to become 
available.

\begin{acknowledgments}
This research is supported in part by the US Department 
of Energy under Contract No. W-7405-ENG-3, 
Grant No. DE-FG02-87ER40371  and
the J. Robert Oppenheimer Fellowship of the 
Los Alamos National Laboratory.
\end{acknowledgments}

%********************************************************************

\begin{widetext}
\vspace*{.5cm}

\begin{appendix}

\section{ Flavor creation and flavor excitation  diagrams}

We emphasize that our need for the matrix elements calculated
here is our interest in matching the $m_c = 0$
charm quarks from the parton distribution functions to the 
$m_c \neq 0$ in the final state.
The internal heavy quark propagators are treated 
as massive. 

\begin{figure}[b!]
\begin{center} 
\psfig{file = 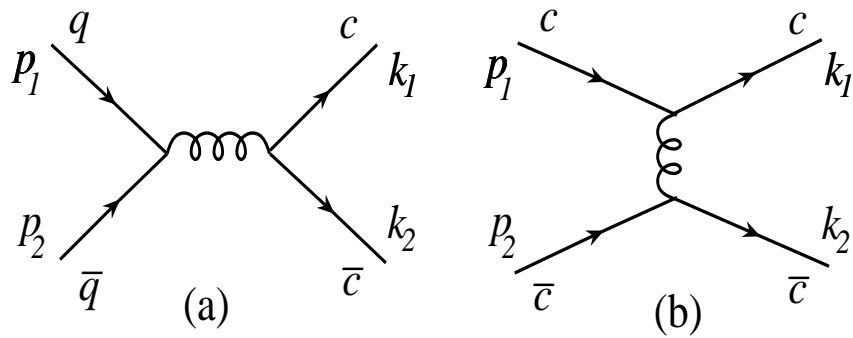,height=2.1in,width=5.in,angle=0}
\caption{(a) Light quark-antiquark annihilation into a massive 
$c \bar{c}$ final state. (b) $t-$channel scattering 
contribution for the case of non-negligible initial (anti)charm
parton distributions. Note that for $c\bar{c} \rightarrow c\bar{c}$
both diagrams contribute. Time increases from left to right in 
all diagrams and 
%%%%%%%%
arrows indicate momentum flow. }  
\label{LightQ}
\end{center} 
\end{figure}

The dominant mode of charm-anticharm di-jet production
and subsequent fragmentation into $D\bar{D}$  mesons 
comes from gluon  fusion and light quark-antiquark  annihilation. 
The $c\bar{c}$ creation diagrams from $q=(u,d,s)$  quarks are shown 
%%%%
in Fig.~\ref{LightQ}(a).  An additional  $t-$channel scattering diagram
contributes  in  Fig.~\ref{LightQ}(b) in the case of non-negligible 
(anti)charm distributions in the nucleon wave-function. 
In all figures the arrows indicate momentum flow. 
The matrix element corresponding to  Fig.~\ref{LightQ}(a) reads:
\begin{eqnarray}  
 i {\cal M}^{(a)}_{q\bar{q}\rightarrow c\bar{c}} & = & 
\bar{u}(k_1)\left( -i g_s \gamma^\mu T^a  \right) {v}(k_2)
\frac{-i g_{\mu \nu} \delta^{a b } }{(p_1+p_2)^2+ i \epsilon} 
\bar{v}(p_2) \left( -i g_s \gamma^\nu T^b  \right) u(p_1) \;.
\label{diagram-Q-s} 
\end{eqnarray}
For identical initial- and final-state quark flavors, the $t-$channel %
must also be considered: 
\begin{eqnarray}  
 i {\cal M}^{(b)}_{c\bar{c}\rightarrow c\bar{c}} &=&  -  
\bar{u}(k_1) \left( -i g_s \gamma^\mu T^a  \right) u(p_1)
\frac{-i g_{\mu \nu} \delta^{a b } }{(p_1-k_1)^2 + i \epsilon} 
\bar{v}(p_2) \left( -i g_s \gamma^\nu T^b  \right) v(k_2) \;.
\label{diagram-Q-t} 
\end{eqnarray}
The relative ``$-$'' between Eqs.~(\ref{diagram-Q-s})  and 
(\ref{diagram-Q-t}) comes from 
Wick's theorem.
\begin{figure}[t!]
\begin{center} 
\psfig{file = 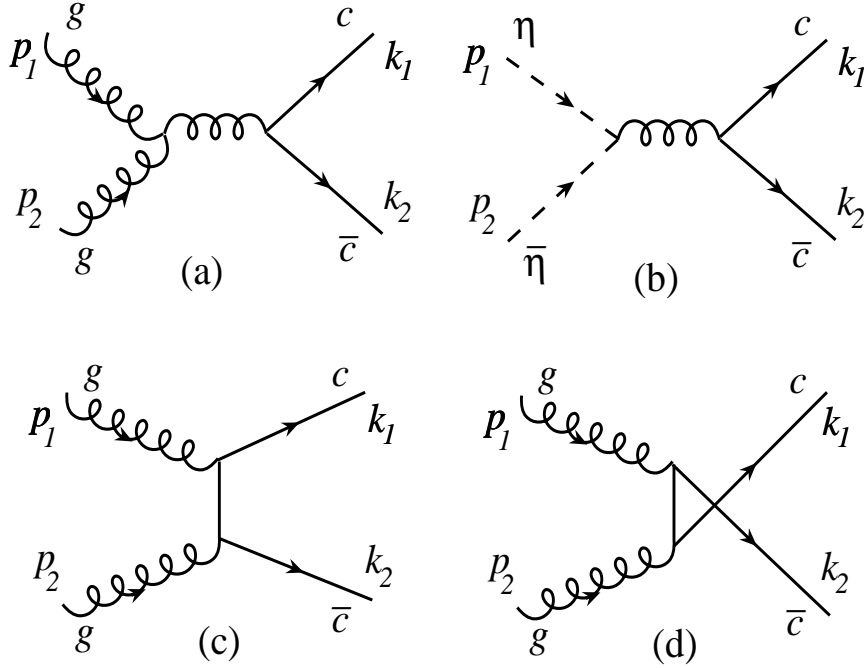,height=3.9in,width=5.in,angle=0}
\caption{(a) Gluon fusion via the $s-$channel  into a 
$c \bar{c}$ final state. (b)  Ghost-antighost annihilation  
that cancels the contribution from the unphysical gluon 
polarization states in the covariant gauge.  
(c) $t-$channel  and (d) the crossed 
$u-$ channel charm exchange diagram.}
\label{Glue}
\end{center} 
\end{figure}
The dominant contribution to the $c\bar{c}$ cross section comes from the
diagrams with  gluonic initial states: %
\begin{eqnarray}  
 i {\cal M}^{(a)}_{gg\rightarrow c\bar{c}} & = &  
\bar{u}(k_1)\left( -i g_s \gamma^\lambda T^c  \right) v(k_2)
\frac{-i g_{\lambda \lambda^\prime}   }{(p_1+p_2)^2 + i \epsilon}  
\nonumber \\
&&  (g_s f^{abc}) \left[ g^{\mu \nu} (p_1 - p_2)^{\lambda^\prime} 
+ g^{\nu \lambda^\prime}(2p_2+p_1)^\mu 
+ g^{ \lambda^\prime \mu} (-2p_1 - p_2)^\nu \right] 
\epsilon_\mu^{\lambda_1}(p_1) \epsilon_\nu^{\lambda_2}(p_2)\;,
\label{diagram-G-s} 
\end{eqnarray}
\begin{eqnarray}  
 i {\cal M}^{(c)}_{gg\rightarrow c\bar{c}} & = & 
\bar{u}(k_1)\left( -i g_s \gamma^\mu T^a  \right) 
i \frac{ \gamma \cdot (k_1-p_1) + m_c  }{(k_1-p_1)^2 - m_c^2 + i \epsilon}  
 \left( -i g_s \gamma^\nu T^b  \right)    v(k_2)
\;  \epsilon_\mu^{\lambda_1}(p_1) \epsilon_\nu^{\lambda_2}(p_2)\;,
\label{diagram-G-t} 
\end{eqnarray}
\begin{eqnarray}  
 i {\cal M}^{(d)}_{gg\rightarrow c\bar{c}} & = & 
\bar{u}(k_1)\left( -i g_s \gamma^\nu T^b  \right) 
i \frac{ \gamma \cdot (p_1-k_2) + m_c  }{(p_1-k_2)^2 - m_c^2 + i \epsilon}  
 \left( -i g_s \gamma^\mu T^a  \right)    v(k_2)
\;  \epsilon_\mu^{\lambda_1}(p_1) \epsilon_\nu^{\lambda_2}(p_2)\;.
\label{diagram-G-u} 
\end{eqnarray}
In the covariant gauge used in this calculation, there are 
contributions at the cross section level coming 
from the unphysical gluon polarizations. The 
diagram ${\cal M}^{(a)}$ in  Fig.~\ref{LightQ}(a) is an 
example of where Fadeev-Popov ghost-antighost annihilation 
diagrams, two of them with $\bar{\eta}$ and $\eta$ interchanged, 
have to be taken into account: 
\begin{eqnarray}  
 i {\cal M}^{(b)}_{\bar{\eta}\eta \rightarrow c\bar{c}} & = &  
\bar{u}(k_1)\left( -i g_s \gamma^\mu  T^b  \right) v(k_2)
\frac{-i g_{\mu \nu}   }{(p_1+p_2)^2 + i \epsilon}  
  (g_s f^{abc}) (-p_1^\mu) \;. 
\label{diagram-Ghost} 
\end{eqnarray}
We recall that a closed ghost loop comes with a ``$-$'' sign and the %
color factor is the same as in $|{\cal M}^{(a)}|^2$. 

\begin{figure}[t!]
\begin{center} 
\psfig{file = 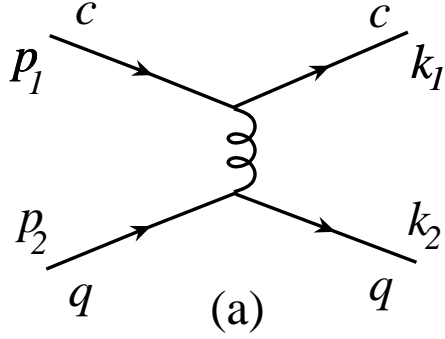,height=2.1in,angle=0}
\caption{(a)  Heavy quark on light quark scattering via $t-$channel gluon 
exchange. 
Diagrams for scattering of anticharm and/or light antiquarks
are similar. }
\label{LightHeavyQ}
\end{center} 
\end{figure}

%%%%%%%%%%%%%%%%%%%%%%%%%%%%%%%%%%%%%%%%%%%%%%%%%%%%%%%%%%%%%%%%%%

For the case of single inclusive $D$ meson production, (anti)charm 
scattering on light quarks and gluons must also be included. We %%%%%
illustrate the calculation for quarks, as shown in 
Fig.~\ref{LightHeavyQ}. Results for antiquarks are
similar. Only one diagram contributes to heavy on light quark 
scattering:
\begin{eqnarray}  
 i {\cal M}^{(a)}_{c q \rightarrow cq} &=&  -  
\bar{u}(k_1) \left( -i g_s \gamma^\mu T^a  \right) u(p_1)
\frac{-i g_{\mu \nu} \delta^{a b } }{(p_1-k_1)^2 + i \epsilon} 
\bar{u}(k_2) \left( -i g_s \gamma^\nu T^b  \right) u(p_2) \;.
\label{diagram-lhQ-t} 
\end{eqnarray}
Figure~\ref{QuarkGlue} shows the contribution of charm quark scattering
on gluons. For $t-$channel gluon exchange, the matrix    %%%%%%%
element, ${\cal M}^{(a)}$, is: %%%%%%%%%
\begin{eqnarray}  
 i {\cal M}^{(a)}_{cg\rightarrow cg} & = &  
\bar{u}(k_1)\left( -i g_s \gamma^\lambda T^c  \right) u(p_1)
\frac{-i g_{\lambda \lambda^\prime}   }{(k_1-p_1)^2 + i \epsilon}  
\nonumber \\
&&  (g_s f^{abc}) \left[ g^{\mu \nu} ( p_2 + k_2 )^{\lambda^\prime} 
+ g^{\nu \lambda^\prime}(p_2 - 2 k_2)^\mu 
+ g^{ \lambda^\prime \mu} (k_2 - 2 p_2)^\nu \right] 
\epsilon_\mu^{\lambda_1}(p_2) \epsilon_\nu^{* \lambda_2}(k_2)\;,
\label{diagram-GC-t} 
\end{eqnarray}
where the unphysical gluon polarizations are again canceled by %
the ghost contribution:
\begin{eqnarray}  
 i {\cal M}^{(b)}_{cg\rightarrow cg} & = &  
\bar{u}(k_1)\left( -i g_s \gamma^\lambda T^c  \right) u(p_1)
\frac{-i g_{\lambda \lambda^\prime}   }{(k_1-p_1)^2 + i \epsilon}  
 (g_s f^{abc}) (-p_2)^{\lambda^\prime}  \;.
\label{diagram-Gghost-t} 
\end{eqnarray}
The other diagrams that  contribute are $u-$channel and %%%
$s-$channel virtual quark exchange, ${\cal M}^{(c)}$ and ${\cal M}^{(d)}$ 
respectively, given for the example of charm by: %%%
\begin{eqnarray}  
 i {\cal M}^{(c)}_{cg\rightarrow cg} & = &  
\bar{u}(k_1)\left( -i g_s \gamma^\mu T^a  \right)
i \frac{ \gamma \cdot (p_1-k_2) + m_c  }{(p_1-k_2)^2 - m_c^2 + i \epsilon}  
\left( -i g_s \gamma^\nu T^b  \right) u(p_1)
\epsilon_\mu^{\lambda_1}(p_2) \epsilon_\nu^{* \lambda_2}(k_2)\;,
\label{diagram-GC-u} 
\end{eqnarray}
\begin{eqnarray}  
 i {\cal M}^{(d)}_{cg\rightarrow cg} & = &  
\bar{u}(k_1)\left( -i g_s \gamma^\nu T^a  \right)
i \frac{ \gamma \cdot (k_1+k_2) + m_c  }{(k_1+k_2)^2 - m_c^2 + i \epsilon}  
\left( -i g_s \gamma^\mu T^b  \right) u(p_1)
\epsilon_\mu^{\lambda_1}(p_2) \epsilon_\nu^{* \lambda_2}(k_2)\;,
\label{diagram-GC-s} 
\end{eqnarray}

\begin{figure}[t!]
\begin{center} 
\psfig{file = 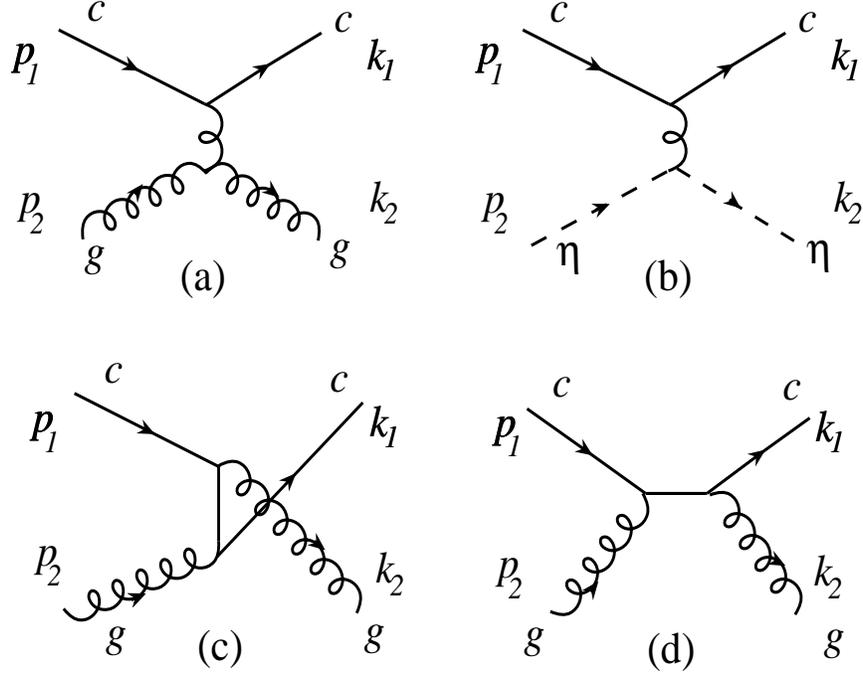,height=3.9in,width=5.in,angle=0}
\caption{(a)  Charm quark on gluon scattering via $t-$channel gluon exchange.  
%%%%
(b)   Charm-ghost scattering 
that cancels the contribution from the unphysical gluon 
polarization states in the covariant gauge.  
(c) Charm quark on gluon scattering via virtual heavy quark 
exchange in the $u-$ channel. 
(d) The corresponding $s-$channel diagram. Anticharm on gluon
scattering is given by a similar set of diagrams. }     %%%
\label{QuarkGlue}
\end{center} 
\end{figure}

Next, we recall the definition of the Mandelstam invariants:
\begin{eqnarray}
\hat{s} &=& (p_1+p_2)^2 = (k_1+k_2)^2 \; , \nonumber \\
\hat{t} &=& (p_1-k_1)^2 = (p_2-k_2)^2 \; , \nonumber \\
\hat{u} &=& (p_1-k_2)^2 = (p_2-k_1)^2 \; . 
\label{mandy}
\end{eqnarray}  
In simplifying the result for the squared matrix elements, 
for charm-anticharm
production we have $\hat{s}+\hat{t}+\hat{u}=2 m_c^2$ and for charm on 
light parton %%
scattering we have $\hat{s}+\hat{t}+\hat{u}= m_c^2$. We have taken 
all massless  initial-state partons.
The spin (polarization) and color averages $\langle \cdots \rangle $ 
are defined in the usual way:
\begin{eqnarray}  
&& \langle \cdots \rangle_{\rm spin} = \frac{1}{2} \sum_{s=\pm 1/2} 
\qquad  {\rm - \; quarks}\;, \qquad 
\langle \cdots \rangle_{\rm polarization} = \frac{1}{2} \sum_{\lambda = 1,2} 
\qquad  {\rm - \; gluons}   \;,  \\
&& \langle \cdots \rangle_{\rm color} = \frac{1}{N_c} 
\sum_{a = 1 \cdots N_c} \qquad  {\rm - \; quarks}\;,  \qquad
 \langle \cdots \rangle_{\rm color} = \frac{1}{(N_c^2-1)} 
\sum_{a = 1 \cdots N_c^2-1} 
\qquad  {\rm - \; gluons}\;,   
\label{averages}
\end{eqnarray}
where $N_c=3$ is the number of colors. Summation over the final-state %%%%%%%%
spin (polarization) and color is understood. %%%%%%

In calculating the color factors
we used factorization of the color algebra and the Lorentz structure in the %%%
Feynman graphs. A diagrammatic approach in non-Abelian gauge theories 
is employed in the calculation below. %%
This is the same approach that was used in the %%%%%%%%
evaluation of the  significantly more involved color 
factors~\cite{Gyulassy:2000er,Gyulassy:2000fs} associated with 
medium-induced gluon bremsstrahlung.
The calculation of the squared amplitudes, 
Eqs.~(\ref{diagram-Q-s})-(\ref{diagram-GC-u}) is straightforward though 
somewhat tedious in the presence of non-zero parton mass
for final-state heavy quarks,
$\sum_{s=\pm 1/2}u^s(p)\bar{u}^s(p) = \gamma \cdot p + m $,    
$\sum_{s=\pm 1/2}v^s(p)\bar{v}^s(p) = \gamma \cdot p - m $.   
In the squared matrix elements given below, % 
the color factors are separated in angle $\langle \;\;\; \rangle$ 
brackets and a ``$-$'' sign inside denotes a destructive color 
interference.  

For light quark  annihilation into a massive  $c\bar{c}$ final state:
\begin{eqnarray}  
\left\langle | {\cal M}_{q\bar{q}\rightarrow c\bar{c}} |^{2} \right\rangle      
%%%%
= \left\langle \frac{2}{9} \right\rangle   \frac{2g_s^4}{\hat{s}^2}
\left( \hat{t}^2 + \hat{u}^2 + 4 m_c^2 \hat{s} - 2 m_c^4    \right)\;.
\label{LightQ-2}
\end{eqnarray}  
For charm quark-antiquark scattering by $t-$channel gluon exchange and 
$s-$channel annihilation
into a massive $c\bar{c}$ pair the result is: 
%%%%%%%%%%%%%%%%%%%%%%%%%%%%%%%%%%%%%%%%%%%%%%%%%%%
\begin{eqnarray}  
\left\langle | {\cal M}_{c\bar{c}\rightarrow c\bar{c}} |^{2} \right\rangle      
%%%%
&=& \left\langle \frac{2}{9} \right\rangle   \frac{2g_s^4}{\hat{s}^2}
\left( \hat{t}^2 + \hat{u}^2 +  4 m_c^2 \hat{s} - 2 m_c^4    \right)
+ \left\langle \frac{2}{9} \right\rangle   \frac{2g_s^4}{\hat{t}^2}
\left( \hat{s}^2 + \hat{u}^2 + 2 m_c^2 \hat{t} - 3 m_c^4    \right) 
\nonumber \\ 
&& + \left\langle - \frac{2}{27} \right\rangle  \frac{4 g_s^4}{\hat{s}\hat{t}}
\left( \hat{u}^2 +  m_c^2 \hat{s} -2 m_c^2 \hat{u} + m_c^4    \right) \;.
\label{HeavyQ-2}
\end{eqnarray}  
Finally, the result for gluon fusion into a heavy charm-anticharm pair, which 
includes %%
the corresponding $t-$channel and $u-$channel scattering, is: %%%%%
\begin{eqnarray}  
\left\langle | {\cal M}_{gg\rightarrow c\bar{c}} |^{2} \right\rangle    %%%%
&=& - \left\langle \frac{3}{16} \right\rangle   \frac{4g_s^4}{\hat{s}^2}
\left( \hat{s}^2  - \hat{t}\hat{u}  +  m_c^2 \hat{s} + m_c^4    \right) 
\nonumber \\  
&&+ \left\langle \frac{1}{12} \right\rangle   \frac{2g_s^4}{(\hat{t}-m_c^2)^2}
\left( \hat{t} \hat{u} + m_c^2 \hat{s}  - 2 m_c^2 \hat{t} - 3 m_c^4  \right) 
+ \left\langle \frac{1}{12} \right\rangle   \frac{2g_s^4}{(\hat{u}-m_c^2)^2}
\left( \hat{t} \hat{u} + m_c^2 \hat{s}  - 2 m_c^2 \hat{u} - 3 m_c^4  \right) 
\nonumber \\  
&& - \left\langle \frac{3}{32} \right\rangle  
\frac{4 g_s^4}{\hat{s}(\hat{t}-m_c^2)}
\left( \hat{t}^2 +  m_c^2 \hat{s} - 2 m_c^2 \hat{t} + m_c^4   \right) 
 + \left\langle - \frac{3}{32} \right\rangle  
\frac{4 g_s^4}{\hat{s}(\hat{u}-m_c^2)}
\left( \hat{u}^2 +  m_c^2 \hat{s} - 2 m_c^2 \hat{u} + m_c^4   \right) 
\nonumber \\  
&&+ \left\langle - \frac{1}{96} \right\rangle  
\frac{4 g_s^4}{(\hat{t}-m_c^2)(\hat{u}-m_c^2)}
\left( m_c^2 \hat{s} - 4 m_c^4   \right) 
\;.
\label{Glue-2}
\end{eqnarray}  

We turn next to the additional contributions to single inclusive charm %%%%
production. The scattering of heavy on light (anti)quarks 
via $t-$channel gluon exchange 
yields: 
\begin{eqnarray}  
\left\langle | {\cal M}_{cq\rightarrow cq} |^{2} \right\rangle  %%%%
&=&  \left\langle \frac{2}{9} \right\rangle   \frac{2g_s^4}{\hat{t}^2}
\left( \hat{s}^2 + \hat{u}^2  +   m_c^2 \hat{t} -  m_c^4    \right). 
\label{HeavyLightQ-1}
\end{eqnarray}  
For the case of charm quarks scattering on gluons, we obtain:
\begin{eqnarray}  
\left\langle | {\cal M}_{cg \rightarrow cg} |^{2} \right\rangle %%%%
&=&  +  \left\langle \frac{1}{2} \right\rangle    
\frac{g_s^4}{ \hat{t}^2 } 
\left( 4  \hat{t}^2  - 4 \hat{s}\hat{u}   -   m_c^2 \hat{s} 
       -  3  m_c^2 \hat{t}   -   m_c^4  \right) 
 \nonumber \\ 
&& -   \left\langle \frac{2}{9} \right\rangle    
   \frac{2 g_s^4}{(\hat{u} - m_c^2)^2} 
\left( \hat{s}\hat{u}  +  2  m_c^2 \hat{u}   -   m_c^2 \hat{s}  \right) 
-   \left\langle \frac{2}{9} \right\rangle    
   \frac{2 g_s^4}{(\hat{s} - m_c^2)^2} 
\left( \hat{s}\hat{u}  +  2  m_c^2 \hat{s}   -   m_c^2 \hat{u}  \right) 
 \nonumber \\ 
&& -  \left\langle - \frac{1}{4} \right\rangle    
  \frac{2 g_s^4}{ \hat{t} (\hat{u} - m_c^2)}
 \left( 2 \hat{u}^2  -  5  m_c^2 \hat{u}   +   m_c^2 \hat{t}  -  m_c^4  \right) 
+  \left\langle  \frac{1}{4} \right\rangle    
  \frac{2 g_s^4}{ \hat{t}   (\hat{s} - m_c^2)}
 \left( 2 \hat{s}^2   -  5  m_c^2 \hat{s}   +   m_c^2 \hat{t}  -  m_c^4  \right) 
 \nonumber \\ 
&& +  \left\langle  - \frac{1}{36} \right\rangle  
  \frac{4 g_s^4}{(\hat{s} - m_c^2) (\hat{u} - m_c^2)}
 \left(  m_c^4  -   m_c^2 \hat{t}   \right) \;.
\label{CharmGlue-1}
\end{eqnarray}  
We can partially check all results by taking the massless, %
$m_c \rightarrow 0$,  limit. If we denote by $(\bar{q})q$ the light %
$u,d,s$ (anti)quarks, we recover the familiar  form %
of the squared matrix elements: 
\begin{eqnarray}  
\lim_{m_c \rightarrow 0} 
\left\langle | {\cal M}_{q\bar{q}\rightarrow c\bar{c}} |^{2}
 \right\rangle      
&=&  \left\langle 
| {\cal M}_{q\bar{q}\rightarrow q^\prime\bar{q}^\prime} |^{2} 
\right\rangle  
\; = \; \frac{4g_s^4}{9} 
\left( \frac{\hat{t}^2 + \hat{u}^2 }{\hat{s}^2} \right) \;, 
\label{LightQ-2-lim}
\end{eqnarray}  
\begin{eqnarray}  
\lim_{m_c \rightarrow 0} 
\left\langle | {\cal M}_{c\bar{c}\rightarrow c\bar{c}} |^{2} \right\rangle      
%%%%
&=&   \left\langle | {\cal M}_{q\bar{q}\rightarrow q\bar{q}} |^{2} \right\rangle        
%%%%
\; = \; \frac{4 g_s^4}{9}
\left(  \frac{\hat{t}^2 + \hat{u}^2}{\hat{s}^2}   + 
   \frac{\hat{s}^2 + \hat{u}^2}{\hat{t}^2}
- \frac{2}{3}   \frac{\hat{u}^2}{\hat{s}\hat{t}}
  \right ) \;,
\label{HeavyQ-2-lim}
\end{eqnarray}  
\begin{eqnarray}  
\lim_{m_c \rightarrow 0} 
\left\langle | {\cal M}_{gg\rightarrow c\bar{c}} |^{2} \right\rangle    %%%%
&=& \left\langle | {\cal M}_{gg\rightarrow q\bar{q}} |^{2} \right\rangle   
\; = \;   \frac{3g_s^4}{8} 
\left( \frac{ \hat{t}^2 + \hat{u}^2 } {\hat{s}^2} \right)
\left( \frac{4}{9} \frac{\hat{s}^2}{\hat{t} \hat{u}} - 1 \right) \;.
\label{Glue-2-lim}
\end{eqnarray}  
\begin{eqnarray}  
\lim_{m_c \rightarrow 0} 
\left\langle | {\cal M}_{cq\rightarrow cq} |^{2} \right\rangle  %%%%
&=&   \left\langle | {\cal M}_{q {q}^\prime \rightarrow q {q}^\prime} |^{2} 
\right\rangle   %%%%
= \frac{4g_s^4}{9}
\left( \frac{\hat{s}^2 + \hat{u}^2}{\hat{t}^2} \right) \;, 
\label{HeavyLightQ-1-lim}
\end{eqnarray} 
\begin{eqnarray}  
\lim_{m_c \rightarrow 0} 
\left\langle | {\cal M}_{cg\rightarrow cg} |^{2} \right\rangle  %%%%
&=&   \left\langle | {\cal M}_{ {q}g  \rightarrow  {q}g } |^{2} \right\rangle   
%%%%
=  g_s^4
\left[  2 \left( 1 - \frac{\hat{s} \hat{u}}{\hat{t}^2 }   \right)  
 -\frac{4}{9}  \left(   \frac{\hat{s}}{\hat{u}}  
            +   \frac{\hat{u}}{\hat{s}}   \right)   - 1 \right] \;.
\label{HeavyQ-G-1-lim}
\end{eqnarray}  
For comparison, see Ref.~\cite{Owens:1986mp}. %
\vspace*{.5cm}

\section{Factorization approach to coherent 
         multiple scattering}

\begin{figure}[b!]
\begin{center} 
\psfig{file = 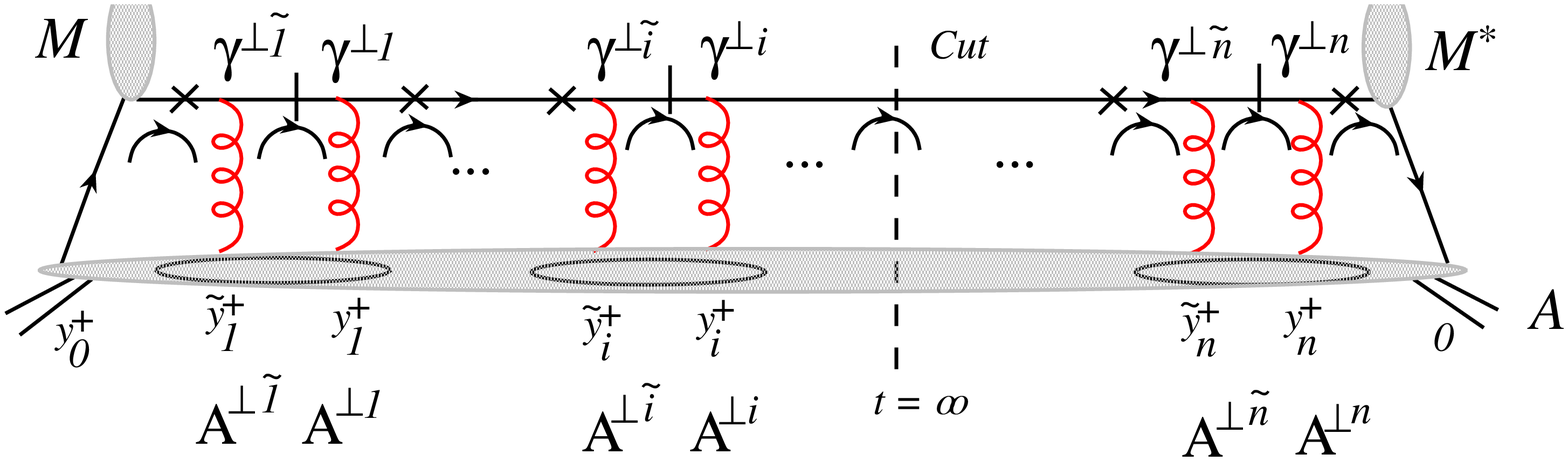,width=6.1in,height=2.in,angle=0}
\caption{Multiple coherent scattering of the outgoing 
partons in proton-nucleus reactions in the $t$-channel. We have 
denoted by $ \longrightarrow  
\hspace*{-0.6cm} \times \hspace*{0.25cm}$ and 
$ \longrightarrow  \hspace*{-0.37cm} \mid \hspace*{0.3cm} $ 
the long distance and contact propagators, respectively. 
We have indicated the lightcone positions, e.g. $y_i^+$.
Arcs show the momentum routing. Gluons are dominated by their 
transverse components and the fields, e.g. $A^{\perp\,i}(y_i^+)$, 
are shown before the conversion to the  field strength tensor 
and before the factorization of the color and Lorentz indexes.} 
\label{tchanel}
\end{center} 
\end{figure}

%%%%%%%%%%%%%%%%%%%%%%%%%%%%%%%%%%%%%%%%%%%%%%%%%

In proton-nucleus reactions at RHIC the negative lightcone direction 
is assigned to the nucleus. Thus, the proton 
has  momentum  $p_A = [p_A^+,0,0_\perp]$ and the nucleus has momentum  
$P_B = [0,P_B^-,0_\perp]$.  We again remind the reader that 
``$B$'' and ``$b$'' are just labels for the target nucleus with 
mass number A. For example,  $p_B \approx P_B / A$ is the 
momentum per nucleon.  Here, we neglect the Fermi momentum and the 
binding energy of the nucleons in the nucleus. For momentum exchange 
$q^\mu = (p_c / z_1 - x_a p_A)^\mu$ we rotate to a frame where
parton ``$d$'' interacts head on with the remnants of the nucleus. 
Let $\bar{n}^\mu = [1,0,0_\perp]$, $n^\mu = [0,1,0_\perp]$  
specify the ``+'' and ``$-$'' lightcone directions, respectively. 
In the equivalent Breit frame
\begin{equation}
q^\mu = - \tilde{x}_b p_B^- n  + 
\frac{-\hat{t}}{2 \tilde{x}_b p_B^-} \bar{n}\;, 
\qquad q^2 = \hat{t} \;. 
\end{equation}
Here, $\tilde{x}_b$ can be derived from the requirement that  
when the  scattered parton ``$d$'' of momentum $x p_B^- n^\mu  + q^\mu$ 
is on shell. The $t=\infty$  cut constrains the momentum fraction from
the nucleus to the known solution,  Eq.~(\ref{xis}),
\begin{equation}
\delta ( (q+x p_B)^2 -m_d^2) \propto \delta(x - x_b)
\qquad \Rightarrow \qquad
 \tilde{x}_b = \frac{x_b}{1+ m_d^2/(-\hat{t})} \;. 
\end{equation}
We note that above, ``$d$'' is the label, not the flavor,
of the outgoing parton. In particular when ``$d$'' is a $c$ 
quark we will have $m_d = m_c = 1.3$~GeV.

Under the standard convention about the ``$+$'' and  ``$-$'' 
lightcone directions at RHIC, it is convenient to use the 
$A \cdot \bar{n} =  A^- = 0$ gauge. 
Such a choice does not
affect any of the physical quantities that appear in the final
result. For example, the parton distributions, the scale of higher
twist and the hard scattering partonic cross sections 
and the Mandelstam variables are all
{\em individually} boost and gauge  invariant.  
We consider the final-state 
interactions of parton ``$d$'' with the  gluons 
from the nucleus with $A^\mu(y_i^+) \approx A^\perp(y_i^+)$  
shown in Fig.~\ref{tchanel}. Using 
\begin{equation}
F^{- \perp}(y_i^+) =   
\frac{\partial A^{\perp}(y_i^+)}{\partial y_i^+} 
- \frac{\partial A^{-}(y_i^+)}{\partial y_i^\perp} 
+ i g \left[ A^{\perp}(y_i^+), A^{-}(y_i^+)  \right]
=  \frac{\partial A^{\perp}(y_i^+)}{\partial y_i^+}  \;,    
\end{equation}
the gluon fields can be represented as components 
of the field strength tensor. The initial-state parton  carries
momentum fraction $x_0$ and  $2n$ gluons carry
momentum fractions $\tilde{x}_i - x_{i-1}$, at positions 
$\tilde{y}^+_i$,
and  fractions $x_{i} - \tilde{x}_i$, at positions ${y}^+_i$,  
of the momenta  $p_B  = P_B/A$ of their corresponding nucleons.
Such choice of the momentum transfers ensures the flow indicated in 
Fig.~\ref{tchanel} and a simple representation of the momentum fraction
in each of the propagators along the scattered parton line.
The part of the diagram associated with the non-perturbative matrix 
element reads:
\begin{eqnarray}
&& 
\int dx_0  \int \frac{d {y}_0^+}{2\pi} \;  
e^{ i x_0  p_B^- {y}_0^+ } \; 
 \left[  \;  \prod_{i=1}^{n} \int  d \tilde{x}_i  \, dx_i  \; 
\int \frac{d \tilde{y}_i^+ p_B^-}{2\pi} \frac{d y_i^+ p_B^- }{2\pi} \; 
e^{ i (\tilde{x}_i - x_{i-1} ) p_B^- \tilde{y}_i^+ }
e^{i( x_i  -  \tilde{x}_i ) p_B^- y_i^+ }\; \right] 
\nonumber \\
&& \qquad \qquad \qquad \qquad \qquad \qquad 
\times  \left \langle P_B \Big| {\cal O}^{init}  
 A^\perp  (y_n^+) A^\perp (\tilde{y}_n^+)  
\cdots  A^\perp  (y_i^+)  A^\perp (\tilde{y}_i^+)  \cdots
A^\perp  (y_1^+)  A^\perp (\tilde{y}_1^+)  
 \Big|  P_B \right \rangle  
%\end{eqnarray}
%\end{document}
\nonumber \\
&& =   \;
\int d x_0  \int \frac{ d {y}_0^+ }{2\pi} \;
e^{i {x}_0  p_B^- {y}_0^+ } \; 
 \left[  \;  \prod_{i=1}^{n} \int d \tilde{x}_i  \, dx_i  \; 
\int \frac{ d\tilde{y}_i^+}{2\pi} \frac{d y_i^+}{2\pi} \; 
\frac{ e^{i (\tilde{x}_i - x_{i-1})p_B^-\tilde{y}_i^+ }}
{i(\tilde{x}_i - x_{i-1} -i\epsilon)}
\frac{ e^{i(  x_{i}  -  \tilde{x}_i )p_B^- y_i^+  }}
{ i ( x_i  -  \tilde{x}_i -i \epsilon ) }\; \; \right] 
\nonumber \\ 
&& \qquad \qquad \qquad \qquad \qquad \qquad 
\times \left \langle P_B \Big|  {\cal O}^{init}  
F^{-\perp}  ({y}_n^+)  F^{-}_{\perp} (\tilde{y}_n^+)  \cdots  
F^{-\perp}  ({y}_i^+)   F^{-}_{\perp} (\tilde{y}_i^+) \cdots    
 F^{-\perp}  ({y}_1^+)  F^{-}_{\perp} (\tilde{y}_1^+) 
\Big|  P_B \right \rangle  \; . \; \;
\label{partial}
\end{eqnarray}
In Eq.~(\ref{partial}) we have carried out an integration by 
parts assuming that the  gluon fields vanish sufficiently fast 
at $|y_i^+| \rightarrow \infty$, $|\tilde{y}_i^+| \rightarrow \infty$. 
The choice of $-i\epsilon $ is convenient for guiding the 
evaluation of the remaining momentum fraction 
integrals~\cite{Qiu:2003vd} but can  be verified by carrying
out  twist-4 calculations in the lightcone and covariant gauges 
and comparing the final result.  
Depending on the type of initial-state parton, the operator
\begin{equation}
{\cal O}^{quark} = \bar{\psi}(0)\frac{\gamma^-}{2} \psi(y_0^+)\;, 
\qquad
{\cal O}^{gluon} =  \frac{1}{x_0 p_B^-}  
F^{-\perp} (0)  F_\perp^-  ({y}_0^+)  \;.
\end{equation}

To factorize the twist-$2+2n$ non-perturbative matrix element
we need to decouple the color and Lorentz indexes, which are 
not shown  explicitly in the equation above. 
We use Fiertz transformations and identify the numerical factors 
associated with the diagonal terms that do not change the
quantum number, for example color, of the parton after a 
pairwise gluon exchange. We obtain $-1/2$ 
for the Lorentz part and 
$1/2N_c=1/6$ or $N_c/(N_c^2-1)=3/8$ for a final-state interacting quark 
or gluon, respectively.  After this factorization of 
indexes, we obtain  pairwise contractions 
$F^{-\perp\,i} (\tilde{y}_i^+)  F_{\perp \,i}^-  ({y}_i^+)$ in the
non-perturbative matrix element, where average over color is implicit,   
and $-\frac{1}{2} (g_{\perp})_{\tilde{i} \,i} \gamma^{\perp\,\tilde{i}}
\cdots \gamma^{\perp\, {i}}$ in the hard scattering matrix element,
which we can evaluate perturbatively in QCD.

Let us now focus on the propagator structure along the outgoing
parton ``$d$''. For definiteness we  consider a heavy quark. 
With {\em net} momentum transfer $x_i p_B^-$ from the nucleus, the 
quark propagator can be represented as follows 
\begin{eqnarray} 
{\cal D} (x_i, x_b,-\hat{t},m_d) &  =  & 
   i\left(\frac{\tilde{x}_b}{-\hat{t}}\right) p_B^- 
\gamma^+  +
i \left(\frac{\tilde{x}_b}{-\hat{t}} \right)
\frac{(\gamma \cdot \tilde{p} + m_d) }
{x_i -x_b + i \epsilon}    \; \;\;\; .
\label{prop}
\end{eqnarray}
In Eq.~(\ref{prop}) the $x_i$-independent vector 
$\tilde{p}$  is on shell
\begin{equation}
\tilde{p}^\mu = (x_b-\tilde{x}_b) p_B^- n^\mu + \frac{ -\hat{t} }
{2 \tilde{x}_b p_B^- } \bar{n}^\mu \;, \qquad 
\tilde{p}^2 = m_d^2 \; .
\end{equation}
We emphasize that the propagators
along the final-state parton line have two distinct 
parts~\cite{Qiu:2003vd,Qiu:2004qk}. The first
one does not have a pole in terms of the yet undetermined momentum 
$x_i p_B^-$. In the Fourier space of the conjugate variable $y^+$, 
the interactions that such a propagator separates can be 
collapsed to a single point, $\propto \delta(\Delta y^+)$.  
Therefore, the short-distance two-gluon exchange can 
be evaluated within the {\em same} nucleon state. 
On the other hand,  the second
part of the propagator has a pole at $x_i = x_b$, which puts the
propagating parton on shell and ensures a long lifetime. 
In Fourier space, the subsequent interactions separated 
by this propagator 
$\propto \theta(\Delta y^+)$ can be long distance and are 
only limited by the nuclear size $\sim r_0 A^{1/3}$, where 
$A$ is the nuclear target mass number. In our notation the 
$t=\infty$ on-shell cut line through parton ``$d$'' reads
\begin{equation}
{\cal C} (x_i, x_b,-\hat{t},m_d) \;  =  \; 
2\pi \left(\frac{\tilde{x}_b}{-\hat{t}}\right) 
( \tilde{p}\cdot \gamma + m_d ) \delta(x_i - x_b) \;.
\label{cut} 
\end{equation}

Naively, the sequence of propagators
separated by the $\gamma^\perp$ matrices may yield an exponentially large 
number of terms per single digram, similar to the one in Fig.~\ref{tchanel}.
However, using the equations of motion for a massive particle, the 
commutation relations for the Dirac algebra and $(\gamma^+)^2 = 
(\gamma^-)^2 = 0$, we can simplify  a typical sequence  of propagators 
along scattered parton ``$d$'' line as follows:
\begin{eqnarray}   
&& \cdots (\tilde{p}\cdot \gamma + m_d) \left[ \frac{g_s^2}{2N_c} 
\left( -\frac{1}{2}
(g_{ \perp})_{i \, \tilde{i} } \right)  
\left(\frac{\tilde{x}_b}{-\hat{t}}\right)^2 
\gamma^{\perp\,i}   
\left(p_B^- \gamma^+  +  \frac{\gamma \cdot \tilde{p} + m_d}
{\tilde{x}_i - x_b + i \epsilon} \right) 
\gamma^{\perp\,\tilde{i}} 
\left(p_B^- \gamma^+  +  \frac{\gamma \cdot \tilde{p} + m_d}
{{x}_{i-1}  - x_b + i \epsilon} \right) 
\right] \cdots \nonumber \\
&=&  \cdots (\tilde{p}\cdot \gamma + m_d) 
\left[  \frac{2 \pi \alpha_s}{3}  
\left(\frac{\tilde{x}_b}{-\hat{t}}\right)  \frac{ \gamma^+ \gamma^-}{2} 
\frac{1}{{x}_{i-1} - x_b + i \epsilon}  \right] \cdots 
\label{simplify}
\end{eqnarray}
where the boundary condition is set by the cut, Eq.~(\ref{cut}), and
we have taken into account that the subsequent scattering has similar
propagator structure. This is the first significant simplification, 
which makes it possible to calculate and resum the multiple final-state
interactions of the struck parton.

The second important simplification comes from a 
model~\cite{Qiu:2003vd} for the high-twist matrix 
elements in Eq.~(\ref{partial}). The natural long and short 
distance separation of soft gluon    %%%
interactions along the trajectory of the struck parton  
justifies the decomposition of the multi-local, multi-parton, %
non-perturbative matrix element in Eq.~(\ref{partial})
in a diagonal basis of individual nucleon states. 
We focus on the QCD aspects of the calculation and not the 
ones related to the nuclear structure. 
To illustrate our idea, we adopt a simplified model of a nucleus of %
constant lab frame number density: 
\begin{equation}
\rho(r)= \frac{A}{V} = \frac{3}{4 \pi r_0^3} \;,  
\label{density}
\end{equation}
where $r_0 = 1.2$~fm is the nucleon radius. 
Each nucleon carries an equal fraction of the momentum of the nucleus, %%%
$p_B \approx P_B/A$, an approximation which is valid so long as the 
energy per nucleon is significantly larger than the binding energy, %
$\epsilon \simeq 8$~MeV, and the momentum per nucleon us significantly 
larger than the Fermi momentum, $k_F\approx 250$~MeV. 
If this condition is violated, the off-diagonal
elements in a density matrix that represent the correlation 
between different nucleons might contribute significantly.  
With the normalization of the  momentum states
$ \langle \, p^\prime \, | \,  p \, \rangle = 2 E_p (2\pi)^3  
\delta^3 ( \vec{p}- \vec{p}^{\, \prime})$,
the decomposition of the matrix element reads:   
\begin{eqnarray}
\left\langle P_B  \Big|  {\cal O}^{init} \prod_{i=1}^n 
F^{-\perp}({y}_i^+) F_\perp^{-}(\tilde{y}_i^+)   \,
 \Big| P_B \right\rangle  & \approx &  
A  \,  \langle p_B | {\cal O}^{init} | p_B  \rangle    
   \prod_{i=1}^n  \frac{\rho(r)}{2 E_{p_B}}   \langle p_B |  
 F^{-\perp}({y}_i^+) F_\perp^{-}(\tilde{y}_i^+)    | p_B  \rangle 
\nonumber \\
&=& A \,   \langle p_B | {\cal O}^{init}    | p_B  \rangle    
   \prod_{i=1}^n  \frac{3}{8 \pi r_0^3 m_N }   \langle p_B |  
  F^{-\perp}({y}_i^+) F_\perp^{-}(\tilde{y}_i^+)  | p_B  \rangle 
\; . \qquad
\label{Matrix-el-dec}
\end{eqnarray}

From Eq.~(\ref{Matrix-el-dec}) it follows that, in the absence of %
final-state coherent interactions, i.e. neglecting the  $\prod_{i=1}^n $
two-gluon correlations,        %    
the minimum bias, inclusive, deeply inelastic  
scattering structure functions, % 
and the minimum bias differential hadron production 
cross section in proton-nucleus reactions, both scale  
with the number of the nucleons, $A$,  
in the nucleus. The expectation value of the bilocal quark 
or gluon operator ${\cal O}^{init}$
enters the definition of the leading-twist 
quark parton distribution function in the 
nucleon and is always present as a first term 
in the twist expansion series.  %%%

Next, we carry out the integrations $\prod_{i=1}^n
\int d \tilde{x}_i d x_i $.
To determine the scale of high-twist corrections per 
nucleon, we recall that the %%
integration over the single pole $\tilde{x}^i$ in Eq.~(\ref{partial}) 
yields 
terms proportional to  $2 \pi\, \theta ( y_i^+ >  \tilde{y}_i^+ )$ 
and  $2 \pi\, \theta ( y_i^+ <  \tilde{y}_i^+ )$. 
If the two gluon exchange is to the left  of the $t = \infty$ cut
we keep the term  $\propto \theta ( y_i^+ >  \tilde{y}_i^+ )$ 
and if  the two gluon exchange is to the right of the 
$t = \infty$ cut, we keep the term 
$ \propto \theta ( y_i^+ <  \tilde{y}_i^+ )$. This ensures that
time grows toward $t = \infty$ and that we keep the dominant
nuclear-size-enhanced contribution. 
Inserting $1 = \exp(i0p_B^- \tilde{y}^+)$ we identify 
\begin{eqnarray}
 (2 \pi)\;  \int \frac{\tilde{y}_i^+}{2\pi}  
 e^{ i0 p_B^- \tilde{y}^+ }  \frac{1}{ p_B^-}
 \langle p_B | F_\perp^{-}(y_i^+)  F^{-\perp}(\tilde{y}_i^+) 
 | p_B  \rangle  \theta ( y_i^+ >\; {\rm or}\;  <  \tilde{y}_i^+ )
 &=&(2\pi)\lim_{x\rightarrow 0}\frac{1}{2} x G(x) \; 
\label{soft-limit}
\end{eqnarray}
by direct comparison with the definition of the parton distribution 
functions~\cite{Qiu:2003vd}. Note that the lightcone positions
$y_i^+,\, \tilde{y}_i^+$ can always be shifted due to 
translational invariance.
The $\theta$-function yields a factor of $1/2$ which 
can  be thought %%
of alternatively  as arising from the symmetry %%%
in the legs of the two-gluon ladder at the interaction point. 
We emphasize that the scale of high-twist corrections depends
on the soft $x \rightarrow 0$ limit of the gluon momentum 
fraction in the nucleon. While finite resolution effects may alter
this result to some small but finite value, 
this is the  value of {\em neither} the Bjorken $x$   %%%%%% 
{\em nor} the momentum fraction fixed by the hard scattering in hadronic
reactions,  Eq.~(\ref{xis}).
As emphasized below Eq.~(\ref{series}), we explicitly keep track    %%%%
of the $A^{n/3}$ enhancement of the multiple soft       %%%%%%%%%%%%%%%%%%%
scattering. We will show below that taking all possible cuts in a 
diagram with $n$ 2-gluon ladders  gives contribution 
$\propto (1/n!) A^{n/3}$. We first identify the prefactors associated 
with such an exponentiating series. A dimensionless equivalent of an average 
path length of a struck parton through the nucleus is given by 
\begin{equation}
  \left\langle 
\int d y_i^+ p_B^- \theta(y_i^+) 
\right\rangle_{\rm Volume}  = \frac{3}{4} m_N r_0 A^{1/3}  \;,
\label{path-length}
\end{equation} 
where the $\langle \cdots \rangle_{\rm Volume}$ is carried out %%%%%
for uniform density, Eq.~(\ref{density}).               %
In a sequence of interactions where the two-gluon ladder contribution
does not vanish, the  $\gamma^+ \gamma^-/2$ matrices give a numerical 
factor of unity; for example, ${\rm Tr}\; \gamma^- (\gamma^+ \gamma^- /2 )^n
\gamma^+ \cdots  =  {\rm Tr}\; \gamma^- \gamma^+ \cdots$.
Collecting all factor from Eqs.~(\ref{simplify}), (\ref{Matrix-el-dec})
and~(\ref{soft-limit}),  we obtain the  parameter
that controls the strength of %
the high-twist corrections on a single nucleon 
(A=1)~\cite{Qiu:2003vd,Qiu:2004qk,Qiu:2004da}
\begin{eqnarray}
\xi^2  & = &  \left( \frac{2 \pi \alpha_s}{3} \right)
\left(    \frac{3}{8 \pi r_0^3 m_N }   \right) 
\left(   \frac{3 m_N r_0  }{4}    \right)
\left( (2\pi) \lim_{x\rightarrow 0}\frac{1}{2} xG(x)     \right) 
\; = \;   \frac{3 \pi \alpha_s}{8\, r_0^2}  
\lim_{x\rightarrow 0}\frac{1}{2} xG(x) \;.
\label{xi2} 
\end{eqnarray}
As we mentioned above, this does not include the $A^{1/3}$
nuclear size  enhancement, nor the derived factor of $1/n!$ which can 
be intuitively understood from path ordered 
interactions~\cite{Gyulassy:2000er}.

\begin{figure}[t!]
\begin{center} 
\psfig{file = 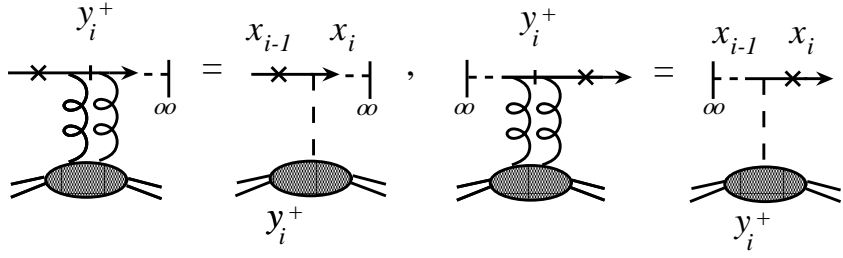 ,width=5.1in,angle=0}
\caption{ Reduced Feynman diagram to the left and right of the 
$t = \infty$ cut. The Feynman rules for the effective scalar 
interaction are given in Eqs.~(\ref{ScalVertL}) and (\ref{ScalVertR}). }
\label{reduced}
\end{center} 
\end{figure}
It is useful to define effective Feynman rules, corresponding to
the diagrams shown in Fig.~\ref{reduced}, for the 
final-state soft scattering of the struck parton:  
\begin{eqnarray}
\label{ScalVertL}
   \tilde{x}_b \left( \frac{\xi^2}{-\hat{t}}  \right) 
\int \frac{d \lambda_i}{2 \pi} \, \frac{e^{i(x_i - x_{i-1})\lambda_i} }
{x_i - x_{i-1} - i\epsilon} 
\frac{ -i  }{x_{i-1} - x_b +i \epsilon }  && \qquad
{\rm left \;\; of} \;\; t=\infty   \;, 
\\[1ex]
   \tilde{x}_b \left( \frac{\xi^2}{-\hat{t}}  \right) 
\int \frac{d \lambda_i}{2 \pi} \, \frac{e^{i(x_i -x_{i-1} )\lambda_i} }
{x_i - x_{i-1} - i\epsilon} 
\frac{ -i  }{x_{i} - x_b -i \epsilon }  && \qquad
{\rm right \;\; of} \;\; t=\infty  \;.   
\label{ScalVertR}
\end{eqnarray}
In Eqs.~(\ref{ScalVertL}) and (\ref{ScalVertR}) we  introduced the notation 
$\lambda_i = p_B^-y_i^+ $.
The difference from our result for DIS, given in Ref.~\cite{Qiu:2003vd}, 
is that we have already performed the density matrix decomposition 
of the high-twist matrix element. Here, $\tilde{x_b}$ replaces
$x_B$ and depends on the heavy quark mass, and  
$Q^2 \rightarrow -\hat{t}$. Numerical factors, including the 
contribution to the non-vanishing traces from the $\gamma$-matrices,
are collected in the scale of higher twist, $\xi^2$, 
Eq.~(\ref{xi2}).

The last part remaining  is to demonstrate that the 
dimensionless integrals can be  performed yielding the 
claimed $A^{1/3}$ amplification of the higher-twist contribution through
the exponentiation of the power series. The twist-$2+2n$ term reads:
\begin{eqnarray}
&& \sum_{i=0}^n  \int d x_0  d x_1  \cdots d x_n   
\int \frac{ d\lambda_0 }{2\pi}  \frac{ d{\lambda}_1 }{2\pi} 
\cdots   \frac{ d {\lambda}_n }{2\pi}  \;
e^{ i x_0 {\lambda}_0 } \;  \frac{1}{p_B^-}  \, 
  \langle p_B | {\cal O}^{init}    | p_B  \rangle  \;  \nonumber \\  
&& \qquad \qquad \qquad  
\times  \prod_{j=1}^i   \frac{e^{i(x_j - x_{j-1})\lambda_j}}
{x_{j-1} - x_b + i\epsilon}
\frac{-i}{x_j - x_{j-1} - i\epsilon }    
   \;   \delta(x_i - x_b)  \prod_{k=i+1}^n 
 \frac{e^{i( x_k - x_{k-1} )\lambda_k}}{x_{k} - x_b - i\epsilon}
\frac{-i}{x_k - x_{k-1} - i\epsilon }  \; .
\label{toughseries}
\end{eqnarray}
In Eq.~(\ref{toughseries}), the appearance of $1/p_B^-$ is associated 
with the conversion of $\tilde{y}_0^+$ to $\lambda_0$.  
There may also be an additional function, $H(x_0)$, which  
contains  smooth $x_0$ dependence. The  $n=0$ first
term corresponds to there being no final-state interactions.

The integrals to the left and right of the $t=\infty$ cut, which 
sets $x_i = x_b$, are taken independently. The integrals over
the $x_j$, $x_k$ poles can be carried out by closing the contour in the 
upper or lower half plane. We note that we have multiple poles.  
These produce $\theta$-functions which  lead path ordering of 
the  $d\lambda_\alpha$ integration  and powers of $\lambda_\alpha - 
\lambda_\beta$.   
Carrying out the $\int d\lambda_1 \cdots  d\lambda_n$,  we find
\begin{eqnarray}
&& 
\int \frac{ d\lambda_0 }{2\pi} e^{ i x_b {\lambda}_0 } \; 
\frac{1}{p_B^-}  \, 
  \langle p_B | {\cal O}^{init}    | p_B  \rangle    \; 
 \sum_{i=0}^n  \frac{i^{i}}{i!}
\left[ - \frac{1}{2}(\Lambda - \lambda_0)^2 \right]^{i}
  \frac{i^{n-i}}{(n-i)!} 
\left[ \frac{1}{2}(\Lambda)^2 \right]^{n-i}
  \;    \nonumber \\
&\approx & 
\int \frac{ d\lambda_0 }{2\pi} e^{ i x_b {\lambda}_0 } \; 
\frac{1}{p_B^-} \, 
  \langle p_B | {\cal O}^{init}    | p_B  \rangle    \; 
 \frac{1}{n!}\,( i \Lambda \lambda_0)^n  \;  =  \; 
 \frac{\Lambda^n}{n!} \frac{d^n}{d x_b^n} \left[ \; 
\int \frac{ d y_0^+ }{2\pi} e^{ i x_b p_B^- y_0^+ } \;
  \langle p_B | {\cal O}^{init}    | p_B  \rangle  \; \right] \;.
\label{toughseries1}
\end{eqnarray}
In Eq.~(\ref{toughseries1}), $\Lambda = A^{1/3} \lambda_0$ and we have
dropped a term ${\cal O}(\lambda_0^2/2)$ which is suppressed by the 
large $A^{1/3}$ nuclear size relative to $\Lambda \lambda_0$.
In the presence of an additional function of $x_0$, for example 
$H(x_0) = |{\cal M}_{ab\rightarrow cd}|^2 / x_0$, the multiple poles
to the left of the $t=\infty$ cut lead to a series of derivatives 
on $H(x_0)$. The result for the modified Eq.~(\ref{toughseries}), 
where we have also included the $(\tilde{x}_b \xi^2/(-\hat{t}))^n$  
power dependence, reads:  
\begin{eqnarray}
&& \sum_{i=0}^n  \left(\tilde{x}_b \frac{\xi^2}{ -\hat{t}}  \right)^n 
\int d x_0  d x_1  \cdots d x_n   
\int \frac{ d\lambda_0 }{2\pi}  \frac{ d{\lambda}_1 }{2\pi} 
\cdots   \frac{ d {\lambda}_n }{2\pi}  \;
e^{ i x_0 {\lambda}_0 } \; \frac{1}{p_B^-} \, 
  \langle p_B | {\cal O}^{init}    | p_B  \rangle    \; 
 H(x_0) \;  \nonumber \\   
&&  \qquad \qquad \qquad   
\times \,\prod_{j=1}^i   \frac{e^{i(x_j - x_{j-1})\lambda_j}}
{x_{j-1} - x_b+i\epsilon}
\frac{-i}{x_j - x_{j-1} - i\epsilon }    % \nonumber \\     
%   && \qquad \qquad \qquad   \qquad \qquad \qquad   \qquad   \qquad 
   \;   \delta(x_i - x_b)  \prod_{k=i+1}^n 
 \frac{e^{i(x_k - x_{k-1})\lambda_k}}{x_{k} - x_b-i\epsilon}
\frac{-i}{x_k - x_{k-1} - i\epsilon }  \;   
\nonumber \\ 
& \approx &     \; \frac{1}{n!}
\left(\tilde{x}_b \frac{\xi^2 A^{1/3}}{ -\hat{t}}  \right)^n 
\frac{d^n}{d x_b^n} \left[ \; H(x_b) 
\int \frac{ d y_0^+ }{2\pi} e^{ i x_b p_B^- y_0^+ } \;
  \langle p_B | {\cal O}^{init}    | p_B  \rangle  \; \right]   \;.
\label{toughseries2}
\end{eqnarray}
Note that all infrared singularities, naively present on the hard 
part, give a finite contribution after taking all possible cuts 
in the diagram, Fig.~\ref{tchanel}.

In studying the modification to the differential cross sections 
due to nuclear-enhanced power corrections, we can either take the scale %
defined in Eq.~(\ref{xi2}) amplified by the nuclear size 
$\propto A^{1/3}$ and a similar effect for the proton with $A=1$, or %
assume that, for the case of the ``elementary'' nucleon, high-twist effects 
have been taken into account in the PDFs and additional interactions %%
in the nucleus are $\propto A^{1/3}-1$. The extraction of the scale of 
power corrections, $\xi^2 = 0.09 - 0.12$~GeV$^2$, for quark scattering, %
from comparison to the shadowing in the nuclear structure function
$F_2^A(x,Q^2)$~\cite{Qiu:2003vd}, was performed using the second %
assumption. Improved analysis may yield somewhat different values for
$\xi^2$. Finally, we recall that in the case of final-state 
gluon scattering, the color singlet coupling, $ N_c / (N_C^2-1) = 3/8 $, 
amplifies the scale of higher-twist effects by a factor of 
$C_A/C_F = 9/4$~\cite{Qiu:2004da}.

To complete the summation of all-twist, nuclear-enhanced power 
corrections, we identify the additional function
$H(x_b) = |M_{ab\rightarrow cd}|^2/x_b$ from Eqs.~(\ref{single})
and~(\ref{double}). In Eq.~(\ref{toughseries}), it multiplies 
the leading-twist  parton distribution $\phi_{b/N}(x_b)$. 
Thus, if we define~\cite{Qiu:2004da} 
$ F_{ab\rightarrow cd}(x_b) = {\phi_{b/N}(x_b) 
|M_{ab\rightarrow cd}|^2}/{x_b}$.
Implementing  the normalization  to  no nuclear effect 
on a single nucleon as follows, $A^{1/3} \rightarrow A^{1/3} -1 $,  
and having allowed for both massless
and massive partons (quarks or gluons) we find:  
\begin{eqnarray}
 F_{ab\rightarrow cd}(x_b) &\Rightarrow & 
\sum_{n=0}^\infty   \frac{1}{n!}
\left(\tilde{x}_b \frac{\xi^2 (A^{1/3}-1)}{ -\hat{t}}  \right)^n 
\frac{d^n}{d x_b^n}  F_{ab\rightarrow cd}(x_b) 
= \exp \left[ \tilde{x}_b \frac{\xi^2 (A^{1/3}-1)}{ -\hat{t}} 
\frac{d}{d x_b}   \right] F_{ab\rightarrow cd}(x_b)
\nonumber \\
 &=& F_{ab\rightarrow cd}\left( x_b + 
 \tilde{x}_b C_d \frac{\xi^2 (A^{1/3}-1)}{ -\hat{t}} \right )  =
F_{ab\rightarrow cd}\left( x_b\left[ 1+ C_d \frac{\xi^2 (A^{1/3}-1)}
{-\hat{t} + m_d^2}  \right] \right)\;.
\label{finalshift}
\end{eqnarray}
In Eq.~(\ref{finalshift}) $C_d = 1\; (9/4)$ for quarks (gluons), 
respectively. When $ m_d \rightarrow 0 $  we recover the result 
of~\cite{Qiu:2004da}.  The mass here reduces the effect of power
corrections and compensates for the fact that $x_b$ already has the 
physical mass of parton ``$d$'' factored in.

\end{appendix}

\hspace*{2.cm}
\end{widetext}

%\vspace*{-.4cm}
%%%%%%%%%%%%%%%%%%%%%%%%%%%%%%%%%%%%%%%%%%%%%%%%%%%%%%%%%%%%%%%%%%%%

\end{document}